\documentclass[prd,amsfonts,onecolumn,superscriptaddress,aps,nofootinbib,11pt]{revtex4-1}

\usepackage[top=3cm,bottom=3cm,left=2cm,right=2cm,marginparwidth=1.75cm]{geometry}

\usepackage{amssymb,amsmath}
\usepackage{subfig}
\usepackage{graphicx}
\usepackage{color}
\usepackage{hyperref}
\usepackage{listings}
\usepackage{soul}
\hypersetup{
    unicode=false,          
    pdftoolbar=true,        
    pdfmenubar=true,        
    pdffitwindow=false,     
    pdfauthor={William},     
    colorlinks=true,       
    linkcolor=blue,          
    citecolor=red,        
    urlcolor=blue
}

\graphicspath{{figures/}}

\providecommand{\be}{ \begin{equation} }
\providecommand{\ee}{\end{equation}}
\providecommand{\bea}{\begin{eqnarray}}
\providecommand{\eea}{\end{eqnarray}}
\providecommand{\nn}{\nonumber}

\def\331{SU(3)_C\otimes SU(3)_L\otimes U(1)_X}
\def\SM{SU(3)_C\otimes SU(2)_L\otimes U(1)_Y}

\begin{document}

\lstset{frame=tb,
  	language=Matlab,
  	aboveskip=3mm,
  	belowskip=3mm,
 	showstringspaces=false,
	columns=flexible,
  	basicstyle={\small\ttfamily},
  	numbers=none,
  	numberstyle=\tiny\color{gray},
 	keywordstyle=\color{blue},
	commentstyle=\color{green},
  	stringstyle=\color{mauve},
  	breaklines=true,
  	breakatwhitespace=true
  	tabsize=3
}

\title{Dynamical Symmetry Breaking and Fermion Mass Hierarchy\\ in the Scale-Invariant 3-3-1 Model\\}
\author{Alex G. Dias}
\email{alex.dias@ufabc.edu.br}
\affiliation{Centro de Ci\^encias Naturais e Humanas, Universidade Federal do ABC,\\
09210-580, Santo Andr\'e-SP, Brasil}
\author{Julio Leite}
\email{julio.leite@ufabc.edu.br}
\affiliation{Centro de Ci\^encias Naturais e Humanas, Universidade Federal do ABC,\\
09210-580, Santo Andr\'e-SP, Brasil}
\affiliation{AHEP Group, Institut de F\'{i}sica Corpuscular --
  C.S.I.C./Universitat de Val\`{e}ncia, Parc Cient\'ific de Paterna.\\
 C/ Catedr\'atico Jos\'e Beltr\'an, 2 E-46980 Paterna (Valencia) - SPAIN}
\author{B. L. S\'anchez-Vega}
\email{bruce@fisica.ufmg.br}
\affiliation{Departamento de F\'isica, UFMG, Belo Horizonte, MG 31270-901, Brasil.}
\author{William C. Vieira}
\email{william.vieira@ufabc.edu.br}
\affiliation{Centro de Ci\^encias Naturais e Humanas, Universidade Federal do ABC,\\
09210-580, Santo Andr\'e-SP, Brasil}
\date{\today}

\begin{abstract}
We propose an extension of the Standard Model (SM) based on the $\331$ (3-3-1) gauge symmetry and scale invariance. Maintaining the main features of the so-called 3-3-1 models, such as the cancellation of gauge anomalies related to the number of chiral fermion generations, this model exhibits a very compact scalar sector. Only two scalar triplets and one singlet are necessary and sufficient to break the symmetries dynamically via the Coleman-Weinberg mechanism. With the introduction of an Abelian discrete symmetry and assuming a natural hierarchy among the vacuum expectation values of the neutral scalar fields, we show that all particles in the model can get phenomenologically consistent masses. In particular, most of the standard fermion masses are generated via a seesaw mechanism involving some extra heavy fermions introduced for consistency. This mechanism provides a partial solution for the fermion mass hierarchy problem in the SM. Furthermore, the simplicity of the scalar sector allows us to analytically find the conditions for the potential stability up to one-loop level and show how they can be easily satisfied. Some of the new particles, such as the scalars $H$, $H^\pm$ and all the non-SM vector bosons, are predicted to get masses around the TeV scale and, therefore, could be produced at the high-luminosity LHC. Finally, we show that the model features a residual symmetry, which leads to the stability of a heavy neutral particle; the latter is expected to show up in experiments as missing energy.

\end{abstract}

\maketitle

\section{Introduction}

The discovery of the Higgs boson \cite{Chatrchyan:2012xdj,Aad:2012tfa}, with a mass $m_h=125.38\pm 0.14$ GeV \cite{Sirunyan:2020xwk}, and the measurements of its main properties \cite{Aad:2015zhl,Khachatryan:2016vau,Sirunyan:2018koj,Aad:2019mbh,Sirunyan:2020xwk}
have shown that the Standard Model (SM) predictions from the spontaneous symmetry breaking 
mechanism -- the Higgs boson couplings to the other SM fields leading to its production cross section 
and branching fractions -- are in agreement with the current experimental observations. It is expected that further data on the Higgs boson properties will improve our understanding about the 
effectiveness of the mechanism of spontaneous symmetry breakdown in the SM and constrain even more the extensions of the SM
containing, in particular, additional scalar bosons. In fact, this has already been done with 
two-Higgs-doublet models and the minimal supersymmetric standard model, for example, but no significant deviation from the SM predictions has been observed so far \cite{Sirunyan:2018koj,Aad:2019mbh}. This can be interpreted as a hint that any successful new high energy theory must have in one of its low energy limits an effective scalar sector that recovers the one in the SM, with one Higgs boson. Nonetheless, a major theoretical drawback of the SM is intrinsically associated with the {\it ad hoc} negative mass term in the scalar potential leading to spontaneous symmetry breaking, which lacks a quantum dynamical origin.

The spontaneous symmetry breaking in the SM is arguably our best understanding of how 
the masses of all the known fermions but neutrinos  arise. It, however, does not provide an explanation for the hierarchy in the value of the fermion masses. For the quarks, we have from the top quark mass $m_t= 172.9\pm 0.4$ GeV to the u-quark mass $m_u=2.16^{+0.49}_{-0.29}$ MeV \cite{Tanabashi:2018oca} a hierarchy of five orders of magnitude. Regarding the leptons, between the mass of the tau $m_\tau=1776.86\pm 0.12$ MeV and the upper bound on the sum of neutrino masses $\sum_\nu m_\nu<0.15$ eV (the lower bound is $\sum_\nu m_\nu>0.06$ eV) \cite{Tanabashi:2018oca}, the hierarchy is even larger spanning ten orders of magnitude at least. Furthermore, the SM cannot account for neutrino oscillation phenomena, once it neither generates small neutrino masses nor large mixing angles (for a review on neutrino physics, see \cite{Mohapatra:2006gs,Strumia:2006db}). This has been one of the main motivations to investigate possible extensions of the SM. 

In this work, we propose a scale-invariant model in which symmetry breaking occurs dynamically according to the Coleman-Weinberg (CW) mechanism \cite{Coleman:1973jx}. The scale invariance implies that no dimensionful parameter is present in the classical Lagrangian so that the tree-level scalar potential contains only quartic terms. Following the dynamical symmetry breaking, a seesaw mechanism takes place leading to a hierarchical mass generation for part of the SM fermions, including neutrinos.
Our theoretical construction is based on a type of 3-3-1 model \cite{Singer:1980sw,Pisano:1991ee,Frampton:1992wt,Montero:1992jk,Foot:1994ym,Hoang:1995vq}, where the $SU(2)_L\otimes U(1)_Y$ symmetry of the SM electroweak sector is extended to the $SU(3)_L\otimes U(1)_X$ symmetry in a particular way which relates the cancellation of gauge anomalies with the number of the observed families of chiral fermions. Different versions of the 3-3-1 models can be classified according to the choice of the $\beta$ parameter defining the electric charge operator in Eq. (\ref{Q}), and we work with a model for which $\beta=1/\sqrt{3}$, however with important differences with respect to the first proposals \cite{Pleitez:1994pu, Ozer:1995xi}. Models invariant under the $SU(3)_L\otimes U(1)_X$ symmetry we consider here have been explored in many contexts, such as that of dark matter \cite{Filippi:2005mt,Mizukoshi:2010ky,Alvares:2012qv,Dong:2013wca,daSilva:2014qba,Montero:2017yvy,Alvarez-Salazar:2019npi,Castellanos:2019kby,Huong:2019vej}, neutrino mass generation and mixing \cite{Dias:2005yh,Catano:2012kw,Dias:2012xp,Pires:2014xsa,Boucenna:2014ela,Fonseca:2016xsy,Das:2020pai}, strong CP problem \cite{Pal:1994ba,Dias:2003iq,Dias:2003zt,Montero:2011tg}, muon anomalous magnetic moment \cite{Kelso:2013zfa,deJesus:2020ngn}, and effects of flavour changing neutral currents \cite{Promberger:2007py,Promberger:2008xg,Cogollo:2012ek,Buras:2012dp,Buras:2013dea,Buras:2016dxz,Queiroz:2016gif}.

The proposed model breaks dynamically both the scale invariance and the $SU(3)_L\otimes U(1)_X$ symmetry down to the $U(1)_Q$ electromagnetic one with a minimal set of scalar fields, two triplets plus a complex singlet, in comparison to typical 3-3-1 models. As a consequence, this minimal scale-invariant 3-3-1 model has a simpler potential and a more compact scalar spectrum. For this simple potential, we establish the stability conditions by imposing the copositive criteria on the matrix of couplings according to the developments in Refs.~\cite{Kannike:2012pe,Kannike:2016fmd,Kannike:2019upf}. 

To study the dynamical symmetry breaking via the CW mechanism, we apply the method of Gildener and Weinberg~\cite{Gildener:1976ih}, which is suitable for obtaining the effective potential in a model with multiple scalar fields. The Gildener-Weinberg method assumes the existence of an energy scale, where the coupling constants are such that there is a flat direction in the tree-level potential. The effective potential, at the one-loop approximation, is then obtained along  this flat direction determining the condition for having a dynamical symmetry breaking. Such a condition requires that the sum of the bosonic field mass to the fourth power times its degrees of freedom must be greater than the corresponding sum for fermionic fields. This fact has been an impediment for the implementation of the CW mechanism in the SM since the dominant contribution from the top quark makes its one-loop effective potential unstable (higher-order corrections can make the effective potential stable but for a Higgs boson mass still incompatible with the experimental value \cite{Elias:2003zm,Chishtie:2010ni}). It has also been observed that the dynamical symmetry breaking of scale-invariant theories can resolve the hierarchy problem since only corrections involving logarithms of the scalar fields are expected for the effective potential \cite{Bardeen:1995kv}, which can be made stable up to the Planck scale in simple extensions of the SM \cite{Meissner:2006zh,  Plascencia:2015xwa, Karam:2015jta, Helmboldt:2016mpi, Khoze:2016zfi, Karam:2016rsz, Brdar:2019qut}.  For a discussion about technical issues of scale invariance and minimal scale-invariant extensions of the SM, see \cite{AlexanderNunneley:2010nw,Helmboldt:2016mpi}.

Each one of the scalar field multiplets of the model is allowed to get a vacuum expectation value (vev) defining, thus, the three energy scales, $v_{\varphi},\,w$ and $v$. The scale $v_\varphi$, coming from the scalar singlet, is assumed to be the largest in the model: $v_\varphi\gg w,v$. The other vevs are due to the scalar triplets and trigger the breaking of the gauge symmetries; $w$ breaks the 3-3-1 gauge symmetry down to the SM group, whereas $v$ is identified with the electroweak scale so that $w\gg v\simeq 246$ GeV. These hierarchies among the energy scales, along with the field content in the model, lead to interesting features in the particle mass spectrum. The model contains just one scalar boson, $h$, at the electroweak scale identified with the discovered $m_h\approx125$ GeV Higgs boson. At the intermediate 3-3-1 breaking scale, $w$, which is assumed here to be around $w\simeq10$ TeV, the model predicts a heavy Higgs boson, $H$, and a charged scalar, $H^\pm$, whose masses could be of few TeV. Completing the scalar particle spectrum, there are two scalar bosons with masses proportional to $v_{\varphi}\simeq 10^3$ TeV, with one of them being the scalon, {\it i.e.} the pseudo-Nambu-Goldstone of the scale invariance breakdown, and the other one a CP-odd scalar which plays a major role in making the one-loop effective potential bounded from below. At this point, it is important to emphasise that the scalar spectrum up to the TeV scale, with only three scalars $h$, $H$ and $H^{\pm}$, is more compact than other popular SM extensions, such as the two-Higgs-doublet model \cite{Branco:2011iw}.

In conventional 3-3-1 models \cite{Singer:1980sw,Montero:1992jk,Foot:1994ym,Hoang:1995vq,Pisano:1991ee,Frampton:1992wt}, it is not possible to generate consistently masses for all the known fermions with a scalar sector containing only two triplets. This happens essentially due to the presence of an accidental chiral symmetry \cite{Montero:2011tg,Montero:2014uya,Sanchez-Vega:2016dwe,Sanchez-Vega:2018qje}. We surpass this problem with the introduction of a set of vector-like fermions that get their dominant mass contribution through their coupling to the complex scalar singlet whose vev is $v_{\varphi}/\sqrt2$. These very heavy fermions, with masses proportional to $v_{\varphi}$, mix with the standard ones, allowing for the implementation of a seesaw mechanism generating masses not only for the active neutrinos but also for most of the known charged fermions. In addition, a hierarchical mass pattern for the standard fermions can be naturally obtained. All these features are more easily noticed with the imposition of a $Z_8$ symmetry, the smallest discrete group for our purposes, on the tree-level scalar potential and the Yukawa Lagrangian. Thus, our model is, in fact, based on the $\331\otimes Z_8$ symmetry group. As a consequence of the $Z_8$ symmetry imposition, an accidental global $U(1)_N$ symmetry arises. This symmetry is broken spontaneously, but there still remains in the model a residual global symmetry, associated with a linear combination of the generators of $SU(3)_L$ and $U(1)_N$, that leads to the stability of the lightest new field which does not mix with the SM ones. We show that, although such a particle cannot, by itself only, explain the observed relic abundance of dark matter in the Universe, it participates in decay processes of the new fermions into SM particles plus missing energy that could be observed at the high luminosity LHC or the future circular collider.

It is worth pointing out that the issue of fermion mass hierarchy and mixing in 3-3-1 models has already been explored by some of us in Refs. \cite{Barreto:2017xix,Dias:2018ddy}. Other interesting solutions to this issue have also been proposed by other authors with the use of discrete symmetries in Refs. \cite{Hernandez:2013hea,Hernandez:2015cra,CarcamoHernandez:2017cwi,CarcamoHernandez:2019iwh} as well as via the Froggatt-Nielsen mechanism in Refs. \cite{Huitu:2017ukq,Huitu:2019kbm,Huitu:2019mdr}. 

This work is organised as follows. In Sec.~\ref{sec:minmod}, we review the 3-3-1 model with two scalar triplets and show that it does not account for a phenomenologically viable fermion spectrum. We then present, in Sec.~\ref{sec:ourmod}, a minimal scale-invariant extension of such a 3-3-1 model, featuring a consistent dynamical symmetry breakdown, which leads to a mechanism of mass generation for all fermions. We study, in Sec.~\ref{scsect}, the scalar sector of the model and derive the stability conditions and the flat direction of the scalar potential. In Sec.~\ref{sec:fermass}, we consider the fermion sector and show the mass generation mechanism which includes a seesaw mechanism for most of the standard fermion masses. Using the results derived in the previous sections, in Sec.~\ref{sec:disc},  the effective potential leading to the dynamical symmetry breaking through the CW mechanism is obtained with the use of Gildener-Weinberg method. In Sec. \ref{sec:resS}, we describe the presence of a residual symmetry and its phenomenological consequences. Finally, our conclusions are presented in Sec.~\ref{sec:conc}. 

\section{An overview of the 3-3-1 model with two scalar triplets}\label{sec:minmod} 

When extending the SM gauge symmetry to $\331$, the SM $SU(2)_L$ doublets need to be embedded into representations of $SU(3)_L$. This can be achieved in an economical manner by embedding each SM doublet into a multiplet transforming in the (anti-)fundamental representation of the extended non-Abelian group. Generically, the 3-3-1 models are defined through its field content and the electric charge operator defined as
\be\label{Q}
    Q = T_3 + \beta \,T_8 + X,  
\ee
where $T_3$ and $T_8$ are the diagonal $SU(3)_L$ generators, and $X$ is the generator of $U(1)_X$. For the current case, we assume $\beta=1/\sqrt{3}$.

For the leptons, the left-handed (LH) fields are arranged into three triplets (one for each family) and the right-handed (RH) charged leptons into $SU(3)_L$ singlets,
\be
 \psi_{i L} =  \left( \nu_i  \ e_i  \  E_i 
\right)_{L}^T \ \sim \left( {\bf 1},{\bf 3}, -2/3 \right), \quad e_{sR} \sim \left({\bf 1},{\bf 1},-1\right),
\label{lep}
\ee
where $i=1,2,3$; $s=1,...,6$; with $e_{4,5,6R}\equiv E_{1,2,3R}$; and the three numbers in parenthesis represent how the fields transform under $SU(3)_C$, $SU(3)_L$ and $U(1)_X$, respectively. Notice that the third component of each triplet is an extra field, $E_{iL}$, and its RH partner, $E_{iR}$, is a $SU(3)_L$ singlet. From Eq. \eqref{Q}, we see that the electric charge of such fields is $q_E = -\frac{1}{3}(2+\sqrt{3}\beta)=-1$, {\it i.e.} the same electric charge as the SM charged leptons.

The quark sector is organised differently. The first two families of the LH quarks are $SU(3)_L$ anti-triplets, while the third transforms as a triplet; the RH quarks are $SU(3)_L$ singlets,
\bea
 Q_{a L} &=&  \left( d_a \  -u_a \ U_{a}\right)_{L}^T \ \sim \left({\bf 3}, {\bf 3^*}, 1/3 \right), \quad Q_{3 L} =  \left( u_3 \  d_3 \ D\right)_{L}^T \ \sim \left({\bf 3},{\bf 3}, 0 \right),\nn\\
 d_{nR}&\sim&\left({\bf 3},{\bf 1},-1/3\right),\quad u_{mR}\sim\left({\bf 3},{\bf 1},2/3\right),\label{qt}
\eea
where $a=1,2$, $n=1,...,4$ and  $m=1,...,5$. We also define the extra quarks as $d_{4}\equiv D$ and $u_{4,5}\equiv U_{1,2}$ carrying the same electric charges as the up-type and down-type quarks, respectively. This unusual  arrangement with two quark families transforming in the anti-fundamental representation is necessary for the cancellation of gauge anomalies in this minimal setup.

When it comes to the scalar sector, at least two triplets are required to perform the expected symmetry breaking which we define as
\be\label{sctri}
\rho=\left(\rho^{0}_1\,\,\rho^{-}_{2}\,\,\rho^{-}_{3}\right)^{T}
 \sim\left(\mathbf{1},\mathbf{3,\,} -2/3\right), \quad \chi=\left(\chi^{+}_{1}\,\,\chi^{0}_{2}\,\,\chi^{0}_{3}\right)^{T}
\sim\left(\mathbf{1},\mathbf{3,\,} 1/3\right).
\ee
The symmetry breaking process can take place spontaneously in two steps. The first step occurs when $\chi_3^0$ acquires a non-vanishing vacuum expectation value (vev), $w/\sqrt{2}$, and the second step takes place through the vev of $\rho_1^0$, $v/\sqrt{2}$ with $w \gg v$,
\be
    \331 \xrightarrow{\langle \chi_3^0 \rangle= w/\sqrt{2}} \SM\xrightarrow{\langle \rho_1^0 \rangle = v/\sqrt{2}} SU(3)_C \otimes U(1)_Q,
\ee
where $U(1)_Q$ is the Abelian group generated by the electric charge operator $Q$, as defined in Eq. (\ref{Q}).
Note that the case where both neutral components of $\chi$ get vev is physically indistinguishable from the current one due to a reparametrisation symmetry connecting the second and third components of the triplet;  see Ref. \cite{Barreto:2017xix} for more details.

The tree-level scalar potential takes the following simple form:
\bea
V(\chi,\rho)=\mu_\rho^2(\rho^\dagger\rho)+\mu_\chi^2(\chi^\dagger\chi)+\lambda_{\rho}(\rho^{\dagger}\rho)^{2}+\lambda_{\chi}(\chi^{\dagger}\chi)^{2} + \lambda_{\rho\chi}(\chi^{\dagger}\chi)(\rho^{\dagger}\rho)+\lambda^\prime _{\rho\chi}(\chi^{\dagger}\rho)(\rho^{\dagger}\chi).
\label{V}
\eea
Its simplicity is also appreciated by noticing that, in addition to the electroweak-scale neutral scalar, $h$, identified with the Higgs boson found at the LHC, the scalar spectrum contains only a heavier CP-even neutral field, $H$, and a heavy charged scalar $H^{\pm}$, with masses given, respectively, by
\be
m_h^2 \simeq \left(2\lambda_\rho - \lambda_{\rho\chi}^2/2\lambda_\chi \right) v^2,\quad m_{H}^2 \simeq  2\lambda_\chi w^2,\quad m_{H^\pm}^2 \simeq \lambda^\prime_{\rho\chi}w^2/2\label{mScm}. 
\ee
Meanwhile, all the remaining scalar degrees of freedom are absorbed in the Higgs mechanism as shown in Ref.~\cite{Barreto:2017xix}.  Therefore, the scalar spectrum is very compact. In fact, it is more compact than in other well-motivated SM extensions, such as left-right \cite{Mohapatra:1974hk,Senjanovic:1975rk,Mohapatra:1974gc,Senjanovic:1978ev,Dias:2019ezk} and two-Higgs-doublet models \cite{Branco:2011iw}.  
If scale invariance is additionally taken into account, the scalar potential in Eq. (\ref{V}) is further simplified, since the terms governed by the dimensionful constants $\mu_\rho$ and $\mu_\chi$ are forbidden. However, without the quadratic terms in the tree-level potential, the calculation of quantum corrections is needed for a clearer understanding of the model as a whole. This will be investigated later in this paper taking into account the symmetry breakdown via the CW mechanism~\cite{Coleman:1973jx}.

\subsection{The gauge sector}\label{subsec:gauge}

As the local gauge group is extended, extra gauge bosons appear. As usual, their masses are obtained from the covariant derivatives acting on the scalar triplets when the scalars acquire vevs. Specifically, from the $(D_{\mu}\rho)^{\dagger}(D^{\mu}\rho)$ and $(D_{\mu}\chi)^{\dagger}(D^{\mu}\chi)$ terms in the Lagrangian, in which the covariant derivative is 
$D^{\mu}=\partial^\mu -igW^\mu_a T^a-ig_X XB^\mu$, 
where the gauge coupling constants of $U(1)_X$ and $SU(3)_L$ groups are related through
the electroweak mixing angle $\theta_W$ according to
\begin{eqnarray}
t^2=\frac{g_X^2}{g^2}=\frac{\textrm{sin}^2\theta_W}{1-\frac{4}{3}\textrm{sin}^2\theta_W},
\label{t2}
\end{eqnarray}
we find that the complex vector bosons,
\begin{eqnarray}
&&W^{\pm}_{\mu} = \frac{W_{1\mu} \mp iW_{2\mu}}{\sqrt{2}},\quad V^{\pm}_{\mu} =\frac{ W_{4\mu} \mp iW_{5\mu}}{\sqrt{2}}, \quad 
V^{0(\dagger)}_\mu =\frac{W_{6\mu}\mp iW_{7\mu}}{\sqrt{2}},
\label{xiv}
\end{eqnarray}
have the following masses:
\begin{eqnarray} m^2_{W^\pm}=\frac{g^2 v^2}{4},\quad m^2_{V^\pm}=\frac{g^2}{4}(v^2+w^2),\quad m^{2}_{V^{0}}= \frac{g^2}{4}w^{2}.
\label{ChVm}
\end{eqnarray}
Furthermore, there are three other vector bosons, the massless photon, $A^\mu$, and two massive neutral  bosons, $Z_1^\mu$ and $Z_2^\mu$, 
\bea 
& & A^{\mu} = \frac{\sqrt{3}}{\sqrt{3+4t^2}}\left( t \  W_{3}^{\mu}+\frac{t}{\sqrt{3}} \ W_{8}^{\mu}+  B^{\mu}\right),\nn\\
& & Z_1^{\mu} = N_{Z_2}\left( -3 m^{2}_{Z_2}\ W_{3}^{\mu}+\sqrt{3} \left(3m^{2}_{Z_2}-g^{2} w^{2} \right)W_{8}^{\mu}+g^{2}w^{2} t \ B^{\mu}\right),\\
& & Z_2^{\mu} =  N_{Z_1}\left(- 3 m^{2}_{Z_1}\ W_{3}^{\mu}+\sqrt{3} \left(3m^{2}_{Z_1}-g^{2} w^{2} \right)W_{8}^{\mu}+g^{2}w^{2} t \ B^{\mu}\right),\nn
\eea 
where
\begin{equation}
N_{Z_2,Z_1}=\left[\left(g^{2}w^{2} t\right)^{2}+\left(3 m^{2}_{Z_2,Z_1}\right)^{2}+3\left(3m^{2}_{Z_2,Z_1}-g^{2}w^{2}\right)^{2}\right]^{-1/2},
\end{equation}
and the approximate masses are given by
\be\label{Zsmass}
m^2_{Z_1}= \frac{g^2v^2}{4\cos^2\theta_W}+{\cal O}\left(\frac{v^2}{w^2}\right),\quad m^2_{Z_2} =\frac{g^2\cos^2\theta_W w^2}{3-4\sin^2\theta_W}+{\cal O}\left(\frac{v^2}{w^2}\right). 
\ee
There are some interesting algebraic relations coming from the symmetry breaking structure of this model which we want to remark. At the tree-level approximation, the vector boson masses satisfy
\begin{equation}
   m^2_{V^\pm}- m^{2}_{V^0}= m^2_{W^\pm} \quad  \textrm{and} \quad
   \frac{m^2_{Z_2}}{m^{2}_{V^0}}  =\frac{\cos^2\theta_W }{\frac{3}{4}-\sin^2\theta_W}+{\cal O}\left(\frac{v^2}{w^2}\right)\approx 1.48\,,
   \label{mrel}
\end{equation}
where we have used ${\rm{sin}}^2\theta_W\simeq0.231$.
\subsection{The Yukawa sector}\label{masslessF}

The attractive features of such an economical 3-3-1 model are, however, not enough 
to make it phenomenologically viable. An important issue is revealed upon the 
derivation of the fermion spectrum. In contrast with experimental evidence, 
some SM fermions remain massless. In the following, we obtain the fermion mass 
matrices and show that this problem has its origins in a global symmetry which appears 
accidentally when the economical setup is considered.

With the fermion and scalar contents presented above, we can write down the Yukawa interactions for leptons and quarks,
\bea  \label{Yuk0}
\mathcal{L}^0_l &=& \overline{\psi_{i L}}\,\chi\left(y^e_{ij}\, e_{j R} + y^E_{ij}\, E_{j R}\right) + \textrm{h.c.},\\
\mathcal{L}^0_q &=& \overline{Q_{a L}}\, \chi^*\left( y^u_{ab}\,u_{b R}+ y^{u_3}_{a3}\,u_{3 R}+ y^U_{ab}\,U_{b R}\right) + \overline{Q_{3 L}}\,\chi\left(y^d_{3b}\, d_{b R}+y^{d_3}_{33}\, d_{3 R}+y^D_{34}\, D_{R}\right)\nn\\
&& + \overline{Q_{a L}}\,\rho^*\left(h^d_{ab}\,d_{b R}+h^{d_3}_{a3}\,d_{3 R}+h^D_{a4}\,D_{R}\right) +
\overline{Q_{3 L}}\,\rho\left(h^u_{3b}\,u_{b R}+h^{u_3}_{33}\,u_{3 R}+h^U_{3b}\,U_{b R}\right) + \textrm{h.c.},\nn
\eea
where the different $y$'s and $h$'s represent the Yukawa coupling matrices.

A straightforward calculation shows that the model has three massless charged leptons and three massless quarks. More specifically, in the basis $(e,E)_{L(R)}^{T}$, we can write the charged-lepton mass matrix as 
\begin{eqnarray}
 M_{\bf E}=\frac{w}{\sqrt{2}}\left(
\begin{array}{c c c}
0 & 0  \\
y^{e} & y^{E} 
\end{array}
\right),
\end{eqnarray}
where all the entries correspond to $3 \times 3$ block matrices. Needless to say, $M_{\bf E}$ has three massless eigenvalues associated with the SM charged leptons which is obviously in disagreement with experimental evidence.

Similarly, in the bases $(d_a, d_3, D)_{L,R}$ and $(u_a, u_3, U_a)_{L,R}$,  we can write the down-type and up-type quark mass matrices, respectively, as
\begin{eqnarray}
M_{\bf D} &=&\frac{1}{\sqrt{2}}
\begin{pmatrix}
h^d_{[2\times2]}\,v &h^{d_3}_{[2\times1]}\, v & h^D_{[2\times1]}\,v  \\
0_{[1\times2]} & 0 & 0 \\ y^d_{[1\times2]}\,w & y^{d_3}\, w & y^D\, w
\end{pmatrix}, \quad M_{\bf U} = \frac{1}{\sqrt{2}}
\begin{pmatrix}
0_{[2\times2]} & 0_{[2\times1]} & 0_{[2\times2]} \\ h^{u}_{[1\times2]}\,v & h^{u_3}\,v & h^U_{[1\times2]}\, v  \\ y^a_{[2\times2]}\,w & y^{u_3}_{[2\times1]}\, w & y^U_{[2\times2]}\, w 
\end{pmatrix},
\end{eqnarray}
and, where not specified, the matrix entry is $1\times1$. From these matrices, we see that 
one down-type and two up-type quarks are massless which brings phenomenological issues.

The presence of massless charged leptons and quarks can be traced back to an accidental 
global symmetry, $U(1)_{PQ}$, which is a Peccei-Quinn like symmetry in the sense that 
it is associated with a $[SU(3)_C]^2\otimes U(1)_{PQ}$ anomaly. As shown in Ref. \cite{Barreto:2017xix}, a residual symmetry associated with $U(1)_{PQ}$ remains
unbroken after spontaneous symmetry breaking. Such an unbroken symmetry is chiral with 
respect to the second components of the fermion (anti-)triplets and their RH singlet 
counterparts, forbidding, in this way, the appearance of mass terms for these fields.
In general, we can see that this is expected in models with minimal scalar sectors in which the families of fermions appear in different representations of the gauge group. Thus, in the present case, when one attempts to reduce the number of scalar triplets to two and takes into account only renormalisable terms in the classical Lagrangian, accidental chiral symmetries arise in the fermion sector. Other 3-3-1 models with similar behaviour can be found in Refs. \cite{Montero:2011tg,Montero:2014uya,Sanchez-Vega:2016dwe}. The issue of fermion masslessness in these models has been solved in Refs. \cite{Ponce:2002sg,Ferreira:2011hm,Dong:2014esa} with the introduction of effective operators.

To generate masses for all fermions, the global $U(1)_{PQ}$ symmetry 
must be broken. This is usually achieved by introducing a third scalar triplet, 
$\eta$, with transformation properties identical to those of $\chi$, but that 
acquires a vev in its second component. Then, it becomes possible 
to generate a mass for all charged fermions, except neutrinos. Nevertheless, 
as in the SM,
neutrino masses and mixings can be generated in a number of ways in 3-3-1 models similar to the one we take into account here \cite{Catano:2012kw, Dias:2012xp, Fonseca:2016xsy}. 
One could simply 
add three right-handed neutrino singlet fields with large Majorana mass terms 
to implement the type-I seesaw  
mechanism~\cite{Minkowski:1977sc,Yanagida:1979as,Mohapatra:1979ia,GellMann:1980vs, Schechter:1980gr},  
as in Ref. \cite{Barreto:2017xix}.  

In the next sections, we present a model extension in which the symmetries 
are broken dynamically via the CW mechanism \cite{Coleman:1973jx}. This is 
done adding a scalar singlet field, a set of vector fermions, three right-handed 
neutrino fields and assuming scale invariance. We will see that besides breaking 
the symmetries in a consistent way, with a scalar potential bounded from below, 
the massless fermions in the model above get masses through a seesaw mechanism.

\section{The Minimal Scale-Invariant 3-3-1 Model}\label{sec:ourmod}

In order to address the phenomenological issue of the massless fermions in the 3-3-1 model with two scalar triplets discussed above, in this section, we propose an extension of the model keeping the scalar sector as simple as possible. As discussed in the previous section, to obtain a consistent mass spectrum for all fermions, the accidental $U(1)_{PQ}$ symmetry must be broken. This is achieved with the introduction of extra fermions instead of the usual extra scalar triplet. The quantum numbers of the extra fermions must allow for operators, in the Yukawa Lagrangian, that break any undesirable accidental chiral symmetry. The main  advantage of this approach is that we preserve all of the appealing features of the effective 3-3-1 model with two scalar triplets, which were discussed above. Moreover, we impose scale invariance on the total Lagrangian. In this way, we further simplify the model since all the dimensionful parameters, such as the arbitrary $\mu$ terms in the scalar potential, are no longer allowed. In the scalar sector, only a complex singlet is added. This field, as we will see, plays important roles in both fermion mass generation and potential stability at quantum level. Another appealing feature is that the fermions left massless in the previous setup, {\it e.g.}, the charged leptons and the bottom quark, become massive through a seesaw-like mechanism. We call the proposed model the {\it minimal scale-invariant 3-3-1 model}. 

In the lepton sector, we introduce
\be\label{newlep}
\Psi_{i L,R} =  \left( \mathcal{E}_{i}^{+}  \ N_{1i}  \  N_{2i}
\right)_{L,R}^T \ \sim \left( {\bf 1},{\bf 3}, 1/3 \right)~, \quad \nu_{iR} \sim ({\bf 1}, {\bf 1}, 0)~, 
\ee
where $\mathcal{E}^{+}$ has an electric charge of $+1$, while $N_1$ and $N_2$ are electrically neutral.
In the quark sector, we add 
\be\label{newquarks}
K_{aL,R} =  \left( \mathcal{A}_{a}^{(5/3)} \  \mathcal{U}_{1a} \ \mathcal{U}_{2a}\right)_{L,R}^T \ \sim \left({\bf 3},{\bf 3}, 1 \right), \quad K_{3L,R} =  \left( \mathcal{B}^{(-4/3)} \  -\mathcal{D}_{1} \ \mathcal{D}_{2}\right)_{L,R}^T \ \sim \left({\bf 3}, {\bf 3^*}, -2/3 \right),
\ee
where $\mathcal{A}^{(5/3)}$ and $\mathcal{B}^{(-4/3)}$ are new quarks with respective electric charges given by the $5/3$ and $-4/3$; whereas $\mathcal{U}$ and $\mathcal{D}$ have the same electric charges as the up-type and down-type quarks, respectively. At last, the scalar sector is extended by one complex singlet
\begin{equation}
\varphi\sim ({\bf 1}, {\bf 1}, 0),
\end{equation}
with $\langle \varphi \rangle = v_\varphi/\sqrt{2}$. The model remains anomaly free since the fermions introduced are either vector-like triplets or gauge singlets. 

In Sec. \ref{sec:fermass}, we will show in detail that all fermions get tree-level masses in this extended model. However, we want to make two remarks in advance. First, the appearance of trilinear operators such as $\overline{\Psi_{R}^{c}}\,\rho^{*}e_{R}$, $\,\overline{\psi_{L}}\,\Psi_{L}^{c}\,\chi^{*}$, $\overline{Q_{L}}\,K_{R}\,\rho$ and $\overline{Q_{3L}}\, K_{3R}\,\rho^*$, explicitly break the accidental Peccei-Quinn like symmetry. Thus, the introduction of the additional fermion fields indeed solves the issue of the massless particles in the 3-3-1 model with two triplets. Second, we impose a $Z_8$ discrete symmetry, under which the fields transform according to Table \ref{tab1}. This discrete symmetry simplifies the spectrum analyses performed in the coming sections by reducing the number of allowed operators in both the scalar potential and Yukawa Lagrangian. 

\begin{table}[h!]
    \centering
\begin{tabular}{|c||cccccccccc|cccccc||cc|c|}
        \hline
  &  $\psi_{iL}$ &
 $ e_{i R}$  &
 $ E_{i R}$ &
 $ \nu_{i R}$  &
 $Q_{a L}$  & 
 $ Q_{3L}$ &
 $u_{iR}$  &
 $U_{aR}$  &
 $d_{iR}$  &  $D_{R}$    &
 $\Psi_{iL}$  &
 $\Psi_{iR}$ &
 $K_{a L}$  &
 $K_{aR}$  &
 $ K_{3L}$  &
  $ K_{3R}$  &
 $\rho$  &
 $\chi$  &
 $\varphi$ \\
 \hline
$Z_8$ & $1$  & $6$ & $0$ & $7$ & $2$ & $3$ & $1$ & $3$ & $4$ & $2$  & $6$ & $4$  & $2$ & $0$ & $3$ & $5$ & $2$ & $1$ & $2$ \\
\hline
\end{tabular} 
 \caption{Field charges under the $Z_8$ symmetry.}
\label{tab1}
\end{table}

\section{Scalar Sector}\label{scsect}

We turn now our attention to the scalar sector composed of two scalar triplets, $\rho$ and $\chi$, and one complex scalar singlet, $\varphi$, which can be written as 
\bea\label{scalars}
\rho^T=\left(\frac{S_1 + i A_1}{\sqrt{2}},\,\rho^{-}_{2},\,\rho^{-}_{3}\right);\,\, \chi^T=\left(\chi^{+}_{1},\,\frac{S_2 + i A_2}{\sqrt{2}},\,\frac{S_3 + i A_3}{\sqrt{2}}\right);\,\,\varphi= \frac{S_\varphi+ i A_\varphi}{\sqrt{2}}.
\eea
With these fields, the most general renormalisable scalar potential, at tree level, is 
\bea\label{VSI_1}
V_0&=& \lambda_{\rho}(\rho^{\dagger}\rho)^{2}+\lambda_{\chi}(\chi^{\dagger}\chi)^{2}  +\lambda _{\rho\chi} \rho^{\dagger}\rho\, \chi^{\dagger}\chi+\lambda^\prime _{\rho\chi}\rho^{\dagger}\chi\,\chi^{\dagger}\rho+\lambda_{\rho\varphi}\rho^{\dagger}\rho\,\varphi^{*}\varphi+\lambda_{\chi\varphi}\chi^{\dagger}\chi\,\varphi^{*}\varphi \nonumber \\
&&+ \lambda_{\varphi}(\varphi^{*}\varphi)^{2} - |\lambda^{\prime}_\varphi|(\varphi^4+\varphi^{*4}).
\eea
The $Z_8$ symmetry in Table  \ref{tab1} simplifies the scalar potential by forbidding non-Hermitian operators, such as $\lambda^\prime_{\rho\varphi}(\rho^\dagger\rho)\varphi^2$. Meanwhile, it allows for the term governed by $|\lambda^{\prime}_\varphi|$ which, as will be shown in Sec. \ref{sec:disc}, is key for the consistency of the model.

The most basic condition that we can impose on the scalar potential couplings comes from the observation that it has to be bounded from below in order to make physical sense. In other words, the vacuum has to be stable. To obtain the constraints associated with such an imposition, it is convenient to rewrite $V_0$ as a biquadratic form of the norm of the fields: $|\rho|,|\chi|,|\varphi|$. More specifically,  let us rewrite Eq. (\ref{VSI_1}) in the compact form $V_0={\bf h}^T\mathbf{\Lambda}(|\theta|,\theta_\varphi){\bf h}$, where ${\bf h}\equiv(|\rho|^{2},|\chi|^{2},|\varphi|^{2})^T\geq0$, and $\mathbf{\Lambda}(|\theta|,\theta_\varphi)$ is the matrix, 
\begin{eqnarray}
\mathbf{\Lambda}(|\theta|,\theta_\varphi)=\left(
\begin{array}{ccc}
 \lambda_\rho  & \frac{1}{2}(\lambda_{\rho \chi} +\lambda^\prime_{\rho \chi}\,
|\theta|) &
   \frac{\lambda \rho \varphi }{2} \\
 \frac{1}{2}(\lambda_{\rho \chi} +\lambda^\prime_{\rho \chi}\,
|\theta|) & \lambda_\chi&
   \frac{\lambda \chi \varphi }{2} \\
 \frac{\lambda \rho \varphi }{2} & \frac{\lambda \chi
   \varphi }{2} &\lambda_\varphi-2|\lambda_\varphi^\prime|\cos(\theta_\varphi)
\end{array}
\right),
\label{lambda}
\end{eqnarray}
where $0\leq|\theta|\leq1$ is the orbit parameter defined as $|\theta|=\hat{\chi}^{*}_i\hat{\rho}_i\hat{\rho}^{*}_j\hat{\chi}_j$, with $i,j=1,2,3$, and $\hat{\chi}_i,\,\hat{\rho}_i=\chi_i/|\chi|,\,\rho_i/|\rho|$. There is another orbit parameter, $\theta_\varphi$, defined as $\varphi=|\varphi|\exp(i\theta_\varphi/4)$. Therefore, the scalar potential, at tree level, is stable if $V_0={\bf h}^T\mathbf{\Lambda}(|\theta|,\theta_\varphi){\bf h}\geq0$. Because $\bf {h}\geq0$, $V_0$ is stable if $\mathbf{\Lambda}(|\theta|,\theta_\varphi)$ is copositive \cite{Kannike:2012pe,Kannike:2016fmd,Sanchez-Vega:2018qje}. 

To find the conditions behind the potential stability, we only need to take into account the values of the orbit space parameters that minimise $V_0$. The fact that $V_0$ is a monotonic function of $|\theta|$ and $\cos \theta_\varphi$ makes our analysis simpler by telling us that the potential reaches its minimum at the boundaries of their respective spaces. As $\cos \theta_\varphi$ appears multiplied by a negative factor, $-2|\lambda^\prime_\varphi|$, the value that minimises the potential is $\cos \theta_\varphi=1$. Whereas for $|\theta|$, the chosen value depends on the sign of $\lambda^\prime_{\rho\chi}$. For $\lambda^\prime_{\rho\chi}>0$, then $|\theta|=0$; otherwise, $|\theta|=1$. We now can apply the copositivity criteria \cite{Kannike:2012pe,Kannike:2016fmd} on $\mathbf{\Lambda}(|\theta|=0,1;\theta_\varphi=0)$ and obtain the inequalities below, which must be simultaneously satisfied by the $\lambda$ couplings,
\bea\label{cop_1}
\lambda_{\rho} &\geq& 0,\quad\quad \lambda_{\chi} \geq 0,\quad\quad \lambda_{\varphi}-2|\lambda_\varphi^\prime| \geq 0,\nonumber \\ 
\overline{\lambda}_1 &\equiv& 2\sqrt{\lambda_{\rho}\lambda_{\chi}}+{\overline{\lambda}}_{\rho\chi}\geq0,\quad \overline{\lambda}_{2}\equiv 2\sqrt{\lambda_{\rho}(\lambda_{\varphi}-2|\lambda_\varphi^\prime|)}+\lambda_{\rho\varphi}\geq0,\quad \overline{\lambda}_{3}\equiv2\sqrt{\lambda_{\chi}(\lambda_{\varphi}-2|\lambda_\varphi^\prime|)}+\lambda_{\chi\varphi}\geq 0,\nn\\
&& \hspace{-.9cm}2\sqrt{\lambda_{\rho}\lambda_{\chi}(\lambda_{\varphi}-2|\lambda_\varphi^\prime|)}+\lambda_{\chi\varphi}\sqrt{\lambda_{\rho}}+\lambda_{\rho\varphi}\sqrt{\lambda_{\chi}}+\overline{\lambda}_{\rho\chi}\sqrt{\lambda_{\varphi}-2|\lambda_\varphi^\prime|}+\sqrt{\overline{\lambda}_{1}\overline{\lambda}_{2}\overline{\lambda}_{3}} \geq 0,
\eea
where $\overline{\lambda}_{\rho\chi}$ takes two values: $\lambda_{\rho\chi}$ and $\lambda_{\rho\chi}+ \lambda^\prime_{\rho\chi}$. 

Let us now look at the symmetry breaking mechanism taking place in the scalar sector and the resulting physical mass spectrum. In principle, due to the scale invariance of the model, the only stationary point of $V_0$ is attained when all neutral scalars are zero. Therefore, one-loop corrections are necessary to shift the tree-level stationary point and, 
in this way, to break spontaneously the gauge symmetries. This is done 
through the Coleman-Weinberg mechanism \cite{Coleman:1973jx}. To implement a 
consistent symmetry breaking mechanism using perturbation methods, we follow the well-known 
Gildener-Weinberg method \cite{Gildener:1976ih}, which generalises the CW mechanism to the case of multiple scalar fields.

The Gildener-Weinberg method relies on
the assumption that, at a given renormalisation scale $\mu_0$, the coupling constants allow for the existence of a direction in the field space along which the potential and its first derivative vanish simultaneously at tree level, known as the flat direction~\cite{Gildener:1976ih}. Nevertheless, the nontrivial degenerate minimum along the flat direction is broken by quantum contributions {\it \`a la} Coleman-Weinberg. Thus, parametrising the scalar fields as $\phi_r\mathbf{N}$, where 
$\phi_r$ is the radial coordinate and $\mathbf{N}$ is a unit vector in the scalar field space, 
we start finding the flat direction, {\it i.e.} the direction in the vacuum surface, 
$\mathbf{N}=\mathbf{n}$, which satisfies: 
i)  $\nabla_{\mathbf{N}}V_0(\mathbf{N})|_{\mathbf{N}=\mathbf{n}}=0$ 
and ii)  $V_0(\mathbf{n})=0$. In addition, the Hessian matrix, $\textrm{P}|_{\mathbf{N}=\mathbf{n}}=\nabla_{\mathbf{N}}\nabla^T_{\mathbf{N}}V_0(\mathbf{N})|_{\mathbf{N}=\mathbf{n}}$, has to be positive semidefinite in order for the flat direction to be a local minimum. 

From the $\nabla_{\mathbf{N}}V_0(\mathbf{N})|_{\mathbf{N}=\mathbf{n}}=0$ condition, we find 
\begin{eqnarray}\label{DofV}
\frac{1}{4}\mathbf{\Lambda}_0.\mathbf{n}^2=\frac{1}{4}\left(
\begin{array}{ccc}
 \lambda_\rho  & \frac{\lambda_{\rho \chi} }{2} &
   \frac{\lambda_{\rho \varphi}}{2} \\
 \frac{\lambda_{\rho \chi} }{2} & \lambda_{\chi} &
   \frac{\lambda_{\chi \varphi}}{2} \\
 \frac{\lambda_{\rho \varphi}}{2} & \frac{\lambda_{\chi
   \varphi}}{2} & \lambda_{\varphi}-2|\lambda_\varphi^\prime| 
\end{array}
\right)\left(
\begin{array}{c}
n_{\rho}^2\\
n_{\chi}^2\\
n_{\varphi}^2
\end{array}
\right)
=\left(
\begin{array}{c}
0\\
0\\
0\end{array}
\right),\label{flatdirectioncondition}
\end{eqnarray}
where ${\mathbf{n}}^T=(n_\rho,n_\chi, n_\varphi)$ is the unit vector in the scalar field space evaluated in the vacuum. We also have that $\mathbf{n}^2$ stands for $(n_\rho^2,n_\chi^2, n_\varphi^2)^T$ and $\mathbf{\Lambda}_0$ is equal to the quartic coupling matrix given in Eq. (\ref{lambda}) with $|\theta|=0$ and $\theta_\varphi=0$, {\it i.e.} $\mathbf{\Lambda}(|\theta|=0,\theta_\varphi=0)$. 

As previously mentioned, Eq. (\ref{flatdirectioncondition}) has, in general, a trivial solution for $\mathbf{n}^2$. In order to find a non-trivial one, the condition
\begin{eqnarray}  
\det \mathbf{\Lambda}_0=\frac{1}{4} \left(4 \lambda_{ \rho}(\lambda_{\varphi}-2|\lambda_\varphi^\prime|)
   \lambda_{\chi}-\lambda_{\rho}\lambda_{\chi \varphi}
   ^2-\lambda_{\rho \varphi}^2 \lambda_{\chi}+\lambda_{\rho \varphi} \lambda_{\rho \chi } \lambda_{
   \chi \varphi}-\lambda_{\rho \chi} ^2 (\lambda_{\varphi}-2|\lambda_\varphi^\prime|)\right)=0 \label{detLambda}
\end{eqnarray}
has to be satisfied \cite{Kannike:2019upf}. This can be seen as if for a given renormalisation scale, $\mu_{0}$, the $\lambda_{\chi \varphi}$ coupling assumes the value  
\begin{eqnarray}
\lambda_{\chi \varphi}|_{\mu_{0}}=\frac{\lambda_{ \rho \varphi}  \lambda_{\rho \chi}
   \pm \sqrt{(\lambda_{\rho \varphi}^2-4 \lambda_{\rho} (
   \lambda_{ \varphi}-2|\lambda_\varphi^\prime|)) (\lambda_{ \rho \chi}^2-4
   \lambda_{\rho} \lambda_{\chi} })}{2 \lambda_{ \rho} }. \label{lambda7}
\end{eqnarray}
Solving Eq. (\ref{flatdirectioncondition}) with $\lambda_{\chi \varphi}|_{\mu_{0}}$ obtained above, $\mathbf{n}^2$ is 
\begin{eqnarray}
n_{\rho}^{2}&=&\frac{-\lambda_{\chi \varphi}  (\lambda_{\rho
   \varphi}+\lambda_{\rho \chi})+2 \lambda_{\chi} 
   (\lambda_{\rho \varphi}-2 (\lambda_{\varphi}-2|\lambda_\varphi^\prime|))+2
   \lambda_{\rho \chi}  (\lambda_{\varphi}-2|\lambda_\varphi^\prime|) +\lambda_{\chi
   \varphi}^2}{\textrm{den}}; \nonumber \\
   n_{\chi}^2&=&\frac{2 \lambda_{\rho}  (\lambda_{ \chi \varphi} -2(
   \lambda_{\varphi}-2|\lambda_\varphi^\prime| ))- \lambda_{\rho \varphi}  (\lambda_{ \rho \chi}+ \lambda_{ \chi \varphi} )+2 \lambda_{ \rho \chi}( \lambda_{\varphi}-2|\lambda_\varphi^\prime|)+\lambda_{ \rho \varphi} ^2}{\textrm{den}};\nonumber \\
   n_{\varphi}^2&=&\frac{2 \lambda_{ \rho}  (\lambda_{ \chi \varphi} -2
   \lambda_{ \chi })-\lambda_{\rho \chi}  (\lambda_{\rho
   \varphi} +\lambda_{\chi \varphi})+2 \lambda_{\rho
   \varphi}  \lambda_{ \chi }+\lambda_{\rho \chi} ^2}{\textrm{den}},\label{nsquared}
\end{eqnarray}
where $\textrm{den}$ is defined as
\begin{eqnarray}
\textrm{den}\equiv&&-4\lambda_\rho (\lambda_\varphi +\lambda_\chi-\lambda_{
   \chi \varphi}-2|\lambda_\varphi^\prime|)-2\lambda_{\rho
   \chi} (\lambda_{\rho \varphi} +\lambda_{\chi \varphi}+4
   |\lambda_\varphi^\prime|)  \nonumber \\
&& -4 \lambda_\chi(-\lambda_{\rho
   \varphi}+\lambda_\varphi-2|\lambda_\varphi^\prime|)   +(\lambda_{\rho \varphi} -\lambda_{\chi \varphi}
   )^2+\lambda_{\rho \chi} ^2+4 \lambda_{\rho \chi} \lambda_\varphi.
\end{eqnarray}
It is also important to note that due to the scale invariance, we have that  $\mathbf{n}\cdot \nabla_{\mathbf{N}}V_0(\mathbf{N})|_{\mathbf{N}=\mathbf{n}}=4V_0(\mathbf{n})$. Therefore, $V_0(\mathbf{n})=0$ for $\mathbf{n}$ given in Eq. \eqref{nsquared}, with $\lambda_{\chi \varphi}$ in Eq. \eqref{lambda7}, which is the ii) condition for the flat direction.

For the solution in Eq. (\ref{nsquared}) to be a local minimum, the Hessian matrix, $\textrm{P}_{ij}$, has to be positive semidefinite on the tangent space of the unit hypersphere at $\mathbf{N}=\mathbf{n}$. More specifically, $\textrm{P}|_{\mathbf{N}=\mathbf{n}}=\mathrm{diag}({\bf{\Lambda}}_0\,\mathbf{n}\circ\mathbf{n})+2{\bf{\Lambda}}_0\circ({\bf{n}}{\bf{n}}^\mathrm{T})$, where $\mathrm{diag}({\bf{\Lambda}}_0\,\mathbf{n}\circ\mathbf{n})$ is the diagonal matrix with its diagonal elements given by ${\bf{\Lambda}}_0\,\mathbf{n}\circ\mathbf{n}$. Also, $(A\circ B)_{ij}=A_{ij}B_{ij}$ stands for the Hadamard product. Taking into account Eq. (\ref{DofV}), it is easy to see that along the flat direction the Hessian matrix is $\textrm{P}|_{\mathbf{N}=\mathbf{n}}=2{\bf{\Lambda}}_0\circ({\bf{n}}{\bf{n}}^\mathrm{T})$. Thus, since $2{\bf{n}}{\bf{n}}^\mathrm{T}$ is manifestly a positive semidefinite matrix and the Hadamard product of two positive semidefinite matrices is also a positive semidefinite matrix, $\textrm{P}_{ij}$ is positive semidefinite if and only if the $\mathbf{\Lambda}_0$ matrix is positive semidefinite. In this case,  
\begin{eqnarray}\label{Hessian}
\textrm{P}&=&\left(
\begin{array}{ccc}
 2 \lambda_\rho  n_\rho^2 & \lambda_{\rho \chi}
  n_\rho n_\chi & \lambda_{\rho \varphi}
   n_\rho n_\varphi \\
 \lambda_{\rho \chi}  n_\rho n_\chi &
   2
   \lambda_\chi n_\chi^2 & \lambda_{\chi \varphi} 
   n_\chi  n_\varphi\\
 \lambda_{\rho \varphi} n_\rho n_\varphi &
   \lambda_{\chi \varphi } n_\chi n_\varphi &
  2 (\lambda_\varphi-2|\lambda_\varphi^\prime|)  n_\varphi^2 \\
\end{array}
\right)
\end{eqnarray}
is positive semidefinite if and only if 
\begin{eqnarray}
&&\lambda_{\rho} \geq 0,\quad\lambda_{\chi} \geq 0,\quad \lambda_{\varphi}-2|\lambda_\varphi^\prime| \geq 0,\quad \mathrm{det}\, {\bf{\Lambda}}_0\geq0, \nonumber \\
-&&2\sqrt{\lambda_{\rho}\lambda_{\chi}}\leq{\lambda}_{\rho\chi}\leq2\sqrt{\lambda_{\rho}\lambda_{\chi}}, \quad -2\sqrt{\lambda_{\rho}(\lambda_{\varphi}-2|\lambda_\varphi^\prime|)}\leq \lambda_{\rho\varphi}\leq 2\sqrt{\lambda_{\rho}(\lambda_{\varphi}-2|\lambda_\varphi^\prime|)}, \nonumber\\ 
-&&2\sqrt{\lambda_{\chi}(\lambda_{\varphi}-2|\lambda_\varphi^\prime|)}\leq\lambda_{\chi\varphi}\leq 2\sqrt{\lambda_{\chi}(\lambda_{\varphi}-2|\lambda_\varphi^\prime|)}\,.\label{semipositive}
\end{eqnarray}
Notice that from Eq. \eqref{detLambda} $\mathrm{det}\, {\bf{\Lambda}}_0=0$ in such a way that the last condition in the first line of the Eq. (\ref{semipositive}) is automatically satisfied.  It is also important to compare conditions coming from the vacuum stability, Eq. \eqref{cop_1}, to the ones coming from positive semidefiniteness of the Hessian matrix $\mathrm{P}$, Eq. \eqref{semipositive}. For  $\lambda_{\rho\chi}^\prime>0$, the conditions in Eq. \eqref{cop_1} are automatically satisfied provided the conditions in Eq. \eqref{semipositive} are. This happens because the positive semidefinite matrices are a subset of the copositive matrices.  However, for $\lambda_{\rho\chi}^\prime< 0$, the matrix that governs the scalar potential behaviour in the limit of large fields is  ${\bf{\Lambda}}(|\theta|=1,\theta_\varphi=0)$ instead of ${\bf{\Lambda}}_0$. Hence, both conditions, Eqs. \eqref{cop_1} and \eqref{semipositive}, must be simultaneously considered.

Once the symmetry breaking pattern at tree level was successfully determined, the scalar mass spectrum can be found. Apart from the would-be Nambu-Goldstone bosons eaten by the gauge fields, in the physical charged sector, there are two mass eigenstates, $H^\pm$, given by
\begin{equation}
H^\pm=\frac{1}{\sqrt{v^2+w^2}}(w\,\rho_3^\pm\,+v\,\chi_1^\pm),
\label{Hpm}
\end{equation}
with a squared mass equal to
\begin{equation}
m_{H^\pm}^2=\frac{\lambda_{\rho
   \chi}^\prime}{2} \left(v^2+w^2\right), \label{masahpm}
\end{equation}
where $v\equiv\sqrt{2}\,n_\rho\langle\phi_r\rangle$, $w\equiv\sqrt{2}\,n_\chi\langle\phi_r\rangle$  and $\langle\phi_r\rangle$ is the breaking scale of scale invariance. From Eq. (\ref{masahpm}), we notice that unless $\lambda^\prime_{\rho\chi}>0$, we would have a tachyonic field. Therefore, the necessary and sufficient conditions for vacuum stability are those shown in Eq. (\ref{semipositive}).

Regarding the CP-even sector, the corresponding mass matrix can be written in terms of the Hessian in Eq. (\ref{Hessian}) as $M_S^2 = \langle\phi_r\rangle$P. Moreover, as previously discussed, $\textrm{P}|_{\mathbf{N}=\mathbf{n}}=2{\bf{\Lambda}}_0\circ({\bf{n}}{\bf{n}}^\mathrm{T})$ and $\mathrm{det}\, {\bf{\Lambda}}_0 =0$, so that $\mathrm{det}\,M_S^2\propto\mathrm{det}\, {\bf{\Lambda}}_0= 0$. This shows that a massless scalar is present in the tree-level spectrum. This massless field is the pseudo-Nambu-Goldstone boson of the scale-invariance symmetry, also known as scalon, defined by 
\begin{equation}\label{scalon}
S=\frac{1}{\sqrt{v^2 + w^2 + v_\varphi^2} }\left[v\,S_1 + w \,S_3 + v_\varphi\,S_\varphi\right]\,.
\end{equation}

The remaining CP-even mass eigenstates, $h$ and $H$, take the following approximate form when the hierarchy $v \ll w \ll v_\varphi$ for the vevs is assumed: 
\begin{eqnarray}
h &\simeq& \frac{1}{N_{h}}\left[S_1 + \frac{\lambda_{\rho \varphi}}{\lambda_{\rho \chi}-\lambda_{\rho \varphi}}\frac{v}{w}S_3 -\frac{v}{v_{\varphi}}S_\varphi\right],\nonumber\\
H &\simeq&\frac{1}{N_{H}}\left[ \frac{\lambda_{\chi}}{\lambda_{\rho \chi}-\lambda_{\rho \varphi}}\frac{v}{w}S_1 + S_3 - \frac{w}{v_{\varphi}}S_\varphi\right], 
\label{h12}
\end{eqnarray}
where $N_{h,H}$ are the normalisation constants.  The exact analytical expressions for the $h$ and $H$ mass eigenstates are omitted here as they are too long and do not bring any essential information at this point. We also have that  $h$ and $H$ have, respectively, the following masses   
\begin{eqnarray}\label{h1h2}
&& m_{h}^2=\lambda_\rho v^2+(\lambda_\varphi-2|\lambda_\varphi^\prime|) v_\varphi^2+\lambda_\chi w^2- m_\Delta^2,\nonumber\\
&& m_{H}^2=\lambda_\rho v^2+(\lambda_\varphi-2|\lambda_\varphi^\prime|) v_\varphi^2+\lambda_\chi w^2+ m_\Delta^2.  
\label{mh12}
\end{eqnarray}
in which 
\begin{eqnarray}
m_\Delta^2= \frac{1}{\lambda_\rho^{1/2}} \,&& \left[\lambda_\rho^3
   v^4+\lambda_\rho  v^2
   \left(v_\varphi^2 \left(\lambda_{\rho \varphi}^2-2 \lambda_\rho (\lambda_\varphi-2|\lambda_\varphi^\prime|)\right)
   +w^2 \left(\lambda_{\rho \chi}^2-2 \lambda_\rho  \lambda_\chi \right)\right)+\lambda_\rho (\lambda_\varphi-2|\lambda_\varphi^\prime|)^2
   v_\varphi^4\right. \nonumber \\ 
&& \left. +v_\varphi^2 w^2 \left(2 \lambda_\rho (\lambda_\varphi-2|\lambda_\varphi^\prime|) \lambda_\chi
   -\lambda_{\rho \varphi}^2 \lambda_\chi+\lambda_{\rho
   \varphi}  \lambda_{\rho \chi}  \lambda_{\chi \varphi}-\lambda_{\rho \chi}^2( \lambda_{\varphi}-2|\lambda_\varphi^\prime|)\right)+\lambda_\rho 
   \lambda_\chi ^2 w^4 \right]^{1/2} \nonumber.
\end{eqnarray}
Assuming the vev hierarchy as well as the minimum conditions given in Eq. (\ref{flatdirectioncondition}), it can be seen that the dominant contributions for the masses are $m_{h}^2 \approx  2 \lambda_\rho v^2$ and $m_{H}^2\approx 2\lambda_{\chi} w^2$. From the previous asymptotic expressions, we can identify $h$ as the SM Higgs of $125$ GeV and, $H$ as an extra scalar with a mass around the 3-3-1 scale. We see from Eq. (\ref{h12}) that the mixing between $h$ and $H$ is small, of the order ${\cal O}(\frac{v}{w})$. Also, under the assumption that $w\ll v_\varphi$, used throughout this work, both CP-even scalar have a small mixing with $S\sim S_\varphi$. The hierarchy of the vevs, with $v_\varphi=\sqrt2 n_\varphi\langle\phi_r\rangle$, implies that in that flat direction  $n_\varphi$ is the dominant component. 

Finally, in the CP-odd sector, there is only one physical eigenstate, $A_\varphi$, with mass equal to 
\begin{equation}\label{massA}
m_{A_\varphi}^2=8|\lambda_\varphi^\prime|v_\varphi^2.
\end{equation}
The pseudoscalar $A_\varphi$ is a component of the gauge singlet $\varphi$, and, as a consequence, it does not have tree-level interactions with the SM particles, except with the Higgs boson. Nonetheless, the interaction with the latter is suppressed by the large mass of $A_\varphi$. As we will see below, $m_{A_\varphi}$ has to be at least of the same order of the vector fermion masses, which along with $A_\varphi$ are supposedly the heaviest states in the model, to ensure the stability the effective potential.  

\section{Fermion spectrum}\label{sec:fermass}

In this section, we analyse the fate of the fermion masses in the minimal scale-invariant 3-3-1 model. We derive the fermion mass matrices and show that all fermions become massive. This procedure is simplified by the $Z_8$ symmetry, presented in Table \ref{tab1}, which restricts the allowed Yukawa interactions, making the mass matrices more manageable. In particular, we show how the fermions that remained massless in the model discussed in Sec. \ref{masslessF} get tree-level masses through a seesaw-like mechanism, when assuming the vev hierarchy: $v\ll w\ll v_\varphi$. Moreover, the results found in the section will allow us to calculate the one-loop effective potential in Sec. \ref{sec:disc}.

\subsection{Lepton masses}

Taking into account all fields and symmetries, we can write down all the renormalisable Yukawa terms involving leptons as
\bea \label{Yuklep}
\mathcal{L}_{l} &=& y^E_{ij}\,\overline{\psi_{i L}}\,\chi E_{j R} + h^{\nu}_{ij}\,\overline{\psi_{iL}}\, \rho\,\nu_{jR} +  h_{ij}^{e}\,\overline{\Psi_{iR}^{c}}\,\rho^{*}e_{jR} + y_{ij}\,\overline{\psi_{iL}}\,\Psi_{jL}^{c}\,\chi^{*}\\
&&+  \frac{f_{ij}^{\nu}}{2}\,\varphi\,\overline{\nu_{iR}^{c}}\,\nu_{jR} + f_{ij}^{\Psi}\,\varphi\,\overline{\Psi_{iL}}\,\Psi_{jR}+ \mbox{h.c.},\nn
\eea 
where $h,\,y$ and $f$ matrices are $3\times3$, and $f^\nu$ can be taken as a $3\times3$ diagonal matrix with real entries without a loss of generality. Furthermore, the term $ y_{ij}\,\overline{\psi_{iL}}\,\Psi_{jL}^{c}\,\chi^{*}$, which contains three $SU(3)_L$ triplets, is implicitly contracted with the totally anti-symmetric tensor $\epsilon_{klm}$ ($k,l,m$ are $SU(3)_L$ indices). For simplicity, we use this convention from here on.

Considering the charged leptons first, we find that $E_L$ and $E_R$ do not mix with the other fields and get the following mass term:
\be\label{Emass}
M_E = \frac{w}{\sqrt{2}} y^E ~,
\ee
where family indices have been omitted. The remaining charged leptons, when grouped in the basis ${\bf \tilde{E}}_{L(R)} \equiv (e,\,\mathcal{E}^{+\,c})^{T}_{L(R)}$, share the $6\times 6$ the mass matrix below,
\begin{eqnarray}\label{Etilmass}
M_{\bf {\tilde E}}=\frac{1}{\sqrt{2}}\left(
\begin{array}{c c c}
0 & -y\,w   \\
h^{e}\,v  & f^{\Psi}\,v_\varphi 
\end{array}
\right)~,
\end{eqnarray}
which is written according to the convention: $\overline{ {\bf{\tilde{ E}}}_{L}}\,  M_{\bf{\tilde{ E}}}\, {\bf {\tilde{E}}}_{R}$. Note that the $Z_8$ symmetry forbids terms like $\overline{\psi_{i L}}\,\chi\, y^e_{ij}\, e_{j R}$, which mix $e_L$ and $E_R$ and would lead to a $9\times9$ mass matrix instead. On the other hand, the vanishing entry in $M_{\bf{\tilde{E}}}$, Eq. (\ref{Etilmass}), does not follow from the $Z_8$ symmetry but the gauge invariance. It is also important to observe that the terms involving $\Psi_{iL,R}$ are essential to solve the issue of the massless fermions present in the original model, thus justifying the introduction of such fields. 

The seesaw-like structure of the mass matrix in Eq. (\ref{Etilmass}) becomes evident when we assume that the vevs are hierarchical. By block diagonalising the squared charged lepton mass matrix $ M_{\bf{\tilde{ E}}}M_{\bf{\tilde{ E}}}^\dagger$, {\it i.e.} writing it as $\textrm{diag}(M_{e^\prime}^2,\,\,M_\mathcal{E^\prime}^2)$, using the methods developed in Refs. \cite{Grimus:2000vj,Hettmansperger:2011bt}, we find
\begin{eqnarray}\label{clMasses}
M_{e^\prime}^2 \simeq \frac{v^2w^2}{2v_{\varphi}^2}y(f^{\Psi})^{-1}h^{e}[y(f^{\Psi})^{-1}h^{e}]^\dagger~\,\,\,\,\,\,\,\,\,\mbox{and}\,\,\,\,\,\,\,\,\,
M_\mathcal{E^\prime}^2 \simeq \frac{v_{\varphi}^2}{2}f^{\Psi}f^{\Psi\dagger}~,
\end{eqnarray}
written, respectively, in the bases
\begin{eqnarray}\label{clStates}
e^\prime_{L}\simeq e_{L}+\frac{w}{v_{\varphi}} y\left(f^{\Psi}\right)^{-1}(\mathcal{E}^{+\,c})_{L}\,\,\,\,\,\,\, \mbox{and}\,\,\,\,\,\,\,\,\,(\mathcal{E^\prime}^{+\,c})_{L} \simeq (\mathcal{E}^{+\,c})_{L} -\frac{w}{v_{\varphi}}\left(f^{\Psi\dagger}\right)^{-1}y^{\dagger}\,e_{L}~.
\end{eqnarray}
Notice that the primed variables are used to distinguish the intermediate states, which are obtained after the block diagonalisation, from the initial flavour states (unprimed). Thus, the primed variables do not yet correspond to the massive physical states, which can only be obtained once one diagonalises the matrices in Eq. (\ref{clMasses}). This convention is used throughout the manuscript. 

From Eq. (\ref{clMasses}), we observe that the largest scale in the model, $v_\varphi$, gives mass to the heavy charged leptons via $M_\mathcal{E^\prime}$ and, at the same time, suppresses the masses of the standard charged leptons in $M_{e^\prime}$ {\it \`a la} the seesaw mechanism. In order to estimate the mass scale of the light leptons, let us assume that $w\simeq10$ TeV and $v_\varphi\simeq 10^3$ TeV. Thus, in $M_{e^\prime}$, the electroweak scale $v=246$ GeV is suppressed by a factor of $w/v_\varphi\simeq \mathcal{O}(10^{-2})$. In this case, the mass of the $\tau$ lepton, the heaviest among the standard leptons, can be naturally  obtained for couplings of order 1. The remaining lepton masses can be fitted by adjusting the relevant Yukawa couplings. Finally, from Eq. (\ref{clStates}), we note that the mixing between standard and non-standard leptons is also suppressed by the factor $w/v_{\varphi}$.

When it comes to the neutral leptons of the model, we can write two independent mass matrices. First, the flavour states $N_{2L}$ and $N_{2R}$ form Dirac fermions, with mass matrix given by
\be\label{N2mass}
M_{N_2} = \frac{v_\varphi}{\sqrt{2}} f^\Psi ~.
\ee 
Second, using the convention $(1/2)\, \overline{ {\bf {\tilde{N} } }_L}\, M_{\bf {\tilde N}}\, ({\bf {\tilde N}}_L)^c$, where ${\bf {\tilde{N}} }_L \equiv \left( \nu_{L},\,\nu _{R}^{c},\, N_{1L},\, N_{1R}^{c}\right)^{T}$, we can write $M_{\bf {\tilde N}}$ as
\begin{eqnarray}\label{NtilM}
M_{\bf{\tilde N}}=\left(
\begin{array}{cc}
 0 & M_{D}^{T} \\
 M_{D} & M_{\varphi} \\
\end{array}
\right),
\end{eqnarray}
with
\begin{eqnarray}\label{NtilM2}
M_{D}=\frac{1}{\sqrt{2}}\left(
\begin{array}{ccc}
 h^{\nu}\,v &  y\,w & 0 \\
\end{array}
\right)^{T}~~~~~~ \mbox{and}~~~~~ M_{\varphi}=\frac{v_{\varphi}}{\sqrt{2}}\left(
\begin{array}{ccc}
f^{\nu} & 0 & 0 \\
0 & 0 & f^{\Psi} \\
0 & f^{\Psi T} & 0 \\
\end{array}
\right)~.
\end{eqnarray}
Although most of the zero entries in these mass matrices are due to the gauge invariance, the $Z_8$ symmetry also plays an important role in simplifying them. For instance, $Z_8$ forbids terms such as $\overline{\Psi_{iL}}\,\chi\,\nu_{jR}$, and, consequently, $N_{2L}$ and $N_{2R}$ do not mix with the other neutral leptons.
The compact structure of the matrix in Eq. (\ref{NtilM}) and the fact that the energy scale of $M_\varphi$ is larger than the $M_D$ one reveal the seesaw structure of such a mass matrix. Upon diagonalising it by blocks, we get
\begin{eqnarray}\label{numass}
M_{\nu^\prime} \simeq -\frac{v^{2}}{\sqrt{2}v_{\varphi}}h^{\nu}(f^{\nu})^{-1}h^{\nu T }\,\,\,\,\,\,\,\,\,\mbox{and}\,\,\,\,\,\,\,\,\,
M_{{\bf N}^\prime}\simeq M_\varphi~,
\end{eqnarray}
written in the bases
\begin{eqnarray}
\nu_{L}^\prime&\simeq&\nu_{L}-\frac{v}{v_{\varphi}}h^{\nu}(f^{\nu})^{-1}\,(\nu_{R})^{c}-\frac{w}{v_{\varphi}}y\left(f^{\Psi T }\right)^{-1} (N_{1R})^{c}\,,\\
{\bf N}^\prime_{L} &\simeq&\left\{(\nu _{R})^{c},\, N_{1L},\, (N_{1R})^{c}\right\} +\left\{\frac{v}{v_{\varphi}}(f^{\nu})^{-1}h^{\nu\dagger},\ 0 \ , \frac{w}{v_{\varphi}}\left(f^{\Psi* }\right)^{-1}y^\dagger \right\}\,\nu_{L}~.\nn
\end{eqnarray}

Note that for $v=246$ GeV and $v_\varphi\simeq 10^3$ TeV, as before, active neutrinos have sub-eV masses for $f^\nu\simeq \mathcal{O}(1)$ and $h^\nu\simeq  \mathcal{O}(10^{-4})$, for example. The remaining neutral leptons have masses around the $v_\varphi$ scale, and similar to the charged lepton case, the mixing angles between active and sterile neutrinos are suppressed by the largest scale in the model, $v_\varphi$, and are, consequently, small.

\subsection{Quark masses}

The quark masses can be obtained from the Yukawa Lagrangian below,
\begin{eqnarray}
 \label{Yukq}
\mathcal{L}_q &=&  y^U_{ab}\,\overline{Q_{a L}}\, \chi^*\,U_{b R} + y^D_{34}\,\overline{Q_{3 L}}\,\chi\, D_{R}+ \overline{Q_{a L}}\,\rho^*\left(h^d_{ab}\,d_{b R}+h^{d_3}_{a3}\,d_{3 R}\right) +
\overline{Q_{3 L}}\,\rho\left( h^u_{3b}\,u_{b R}+h^{u_3}_{33}\,u_{3 R} \right)\nn\\ &&+ 
\tilde{h}_{ab}\,\overline{Q_{aL}}\,K_{bR}\,\rho + h_{33}\,\overline{Q_{3L}}\, K_{3R}\,\rho^*+\overline{K_{aL}}\,\chi\left(\tilde{y}^u_{ab}\,u_{bR}+\tilde{y}^{u_3}_{a3}\,u_{3R}\right)  + \overline{K_{3L}}\,\chi^*\left(\tilde{y}^d_{3b}\,d_{bR}+\tilde{y}^{d_3}_{33}\,d_{3R}\right)\nn\\
&& +  f^{K_a}_{ab}\,\varphi\, \overline{K_{aL}}\, K_{b R} + f^{K_3}_{33}\,\varphi^*\, \overline{K_{3L}}\,K_{3R} + \mbox{h.c.}\,
\end{eqnarray}
First, we consider the up-type quarks for which we obtain two independent mass matrices. If we choose as bases: $\mathbf{U}_{L,R}^{(1)} \equiv (u_a, u_3, \mathcal{U}_{2a})_{L,R}$ and $\mathbf{U}_{L,R}^{(2)} \equiv (U_a, \mathcal{U}_{1a})_{L,R}$, we can write a $5\times5$ and a $4\times 4$ mass matrix,
\begin{eqnarray}\label{newUmass}
M_{\bf U}^{(1)} =\frac{1}{\sqrt{2}}
\begin{pmatrix}
0_{[2\times2]} & 0_{[2\times1]} & - \tilde{h}_{[2\times2]}\,v \\ h^u_{[1\times2]}\,v & h^{u_3}_{[1\times1]}\,v & 0_{[1\times2]}\, \\ \tilde{y}^u_{[2\times2]}\,w & \tilde{y}^{u_3}_{[2\times1]}\, w & f^{K_a}_{[2\times2]}\, v_\varphi
\end{pmatrix}~ \quad\mbox{and}\quad
M_{\bf U}^{(2)} = \frac{1}{\sqrt{2}}
\begin{pmatrix}
y^U_{[2\times2]}\,w & -\tilde{h}_{[2\times2]}\,v \\ 0_{[2\times2]} & f^{K_a}_{[2\times2]}\, v_\varphi
\end{pmatrix}.
\end{eqnarray}
Similar to the lepton sector, the $Z_8$ symmetry simplifies the mass matrices in the quark sector. For example, the terms $\overline{Q_{a L}}\, \chi^* y^u_{ab}\,u_{b R}$ and $\overline{Q_{3 L}}\,\rho\,h^U_{3b}\,U_{b R}$ are not allowed by $Z_8$, and the mass matrices $M_{\bf U}^{(1)}$ and $M_{\bf U}^{(2)}$ become independent. Furthermore, we must emphasise the importance of the extra quark triplets $K_{b L,R}$ in solving the masslessness problem in the up-type quark sector. The introduction of $K_{bR}$ allows for the term $\tilde{h}_{ab}\,\overline{Q_{aL}}\,K_{bR}\,\rho$, which mixes $u_{aL}$, originally massless, and $\mathcal{U}_{aR}$. Meanwhile, the presence of $K_{aL}$, in addition to contributing to the cancellation of anomalies, allows for the term $f^{K_a}_{ab}\,\varphi\, \overline{K_{aL}}\, K_{b R}$ which provides a large mass scale for $M_{\bf U}^{(1)}$, leading to a seesaw mechanism, as described below.

The matrix $M_{\bf U}^{(1)}$ contains three light quarks mixed with two heavy quarks. We block diagonalise $M_{\bf U}^{(1)}M_{\bf U}^{(1)\dagger}$ by rotating the left-hand fermions to separate the ordinary from the exotic quarks and find
\begin{eqnarray}\label{SMumm}
M_{{\bf u}^\prime}^2 &=&\frac{1}{2}
\begin{pmatrix}
\left(\frac{vw}{v_\varphi}\right)^2   \tilde{h} (f^{K_a})^{-1} (Y^u)^2 (f^{K_a \dagger})^{-1} \tilde{h}^\dagger & \frac{v^2w}{v_\varphi}  h (f^{K_a})^{-1}\left( \tilde{y}^u h^{u\dagger} +\tilde{y}^{u_3} h^{u_3 *}\right) \\ \star^\dagger &   v^2 \, A_u^2
\end{pmatrix},\\ 
M_{{\bf U}^\prime}^2 &=&  \frac{v_\varphi^2}{2}
f^{K_a} f^{K_a\dagger},\nonumber
\end{eqnarray}
with $(Y^u)^2=\tilde{y}^u\tilde{y}^{u\dagger} + \tilde{y}^{u_3} \tilde{y}^{u_3 *}$ and $A_u^2=h^u h^{u\dagger}+h^{u_3} h^{u_3*}$, where $M_{{\bf u}^\prime}^2$ is written in the basis
\bea\label{ubasis}
u^\prime_{aL}&=& u_{aL} +\frac{v}{v_\varphi}\left[\tilde{h}(f^{K_a})^{-1}\right]_{ab}\mathcal{U}_{2bL}\,, \\ u^\prime_{3L} &=& u_{3L} -\frac{vw}{v_\varphi^2}
\left(h^u\tilde{y}^{u\dagger}+h^{u_3}\tilde{y}^{u_3\dagger}\right)_{a}  \left(f^{K_a} f^{K_a\dagger}\right)_{ab}^{-1}\mathcal{U}_{2bL}\,,\nn
\eea
whereas the basis for $M_{{\bf U}^\prime}^2$ is 
\be\label{Ubasis}
\mathcal{U}^\prime_{2aL} = \mathcal{U}_{2aL} -\frac{v}{v_\varphi}\left[(f^{K_a\dagger})^{-1} \tilde{h}^\dagger\right]_{ab} u_{bL} + \frac{vw}{v_\varphi^2} \left(f^{K_a} f^{K_a\dagger}\right)^{-1}_{ab}
\left(\tilde{y}^u h^{u\dagger}+\tilde{y}^{u_3} h^{u_3*}\right)_b  u_{3L} \,.
\ee
For simplicity, the sizes of the Yukawa matrices, originally shown in Eq. (\ref{newUmass}), have been omitted in Eqs. (\ref{SMumm}), (\ref{ubasis}) and (\ref{Ubasis}).

From $M_{{\bf u}^\prime}^2$ in Eq. (\ref{SMumm}), we can see that while the third family gets a mass proportional to the electroweak scale $v$, the first two families get masses proportional to $(w/v_\varphi) v \ll v$ due to a seesaw-like mechanism  that takes place as a result of the mixing with the heavy up-type quarks. In this way, a mass hierarchy between the third and the other families is present.

The other matrix in Eq. (\ref{newUmass}), $M_{\bf U}^{(2)}$, is approximately diagonal as the off-diagonal terms are much smaller than the diagonal ones. From it, we obtain two heavy up-type quarks with masses proportional to $w$, and the other two are even heavier with masses proportional to $v_\varphi$, while the mixing angles are very suppressed.

With the down-type quarks, we find a similar situation. The corresponding fields can be grouped into two independent bases: $\mathbf{D}_{L,R}^{(1)} \equiv (d_a, d_3, \mathcal{D}_{2})_{L,R}$ and $\mathbf{D}_{L,R}^{(2)} \equiv (D, \mathcal{D}_{1})_{L,R}$, according to which the respective $4\times4$ and $2\times2$ mass matrices can be written
\begin{eqnarray}\label{newDmass}
M_{\bf D}^{(1)} =\frac{1}{\sqrt{2}}
\begin{pmatrix}
h^d_{[2\times2]}\,v & h^{d_3}_{[2\times1]}\, v & 0_{[2\times1]}\\ 0_{[1\times2]} & 0_{[1\times1]} & h_{[1\times1]}\,v \\  
\tilde{y}^d_{[1\times2]}\,w  & \tilde{y}^{d_3}_{[1\times1]}\, w & f^{K_3}_{[1\times1]}\,v_\varphi
\end{pmatrix}\quad\mbox{and}\quad
 M_{\bf D}^{(2)} =  \frac{1}{\sqrt{2}}
\begin{pmatrix}
 y^D_{[1\times1]}\, w & h_{[1\times1]}\,v \\
 0_{[1\times1]} & f^{K_3}_{[1\times1]}\,v_\varphi
\end{pmatrix}.
\end{eqnarray}
Once again, the $Z_8$ symmetry simplifies the mass matrices, and, here, it makes $M_{\bf D}^{(1)}$ and $M_{\bf D}^{(2)}$ independent. Moreover, the introduction of $K_{3L,R}$ allows for the appearance of the necessary terms to make all the down-type quarks massive, {\it e.g.} $h_{33}\,\overline{Q_{3L}}\, K_{3R}\,\rho^*$ and $f^{K_3}_{33}\,\varphi^*\, \overline{K_{3L}}\,K_{3R}$.

Upon block diagonalisation of $M_{\bf D}^{(1)}(M_{\bf D}^{(1)})^\dagger$, we find
\begin{eqnarray}\label{SMqmm}
M_{{\bf d}^\prime}^2 &=&\frac{1}{2}
\begin{pmatrix}
v^2\,(A^d)^2 & -\frac{v^2w}{v_\varphi} \left(h^d\tilde{y}^{d\dagger}+h^{d_3}\tilde{y}^{d_3*} \right) (f^{K_3*})^{-1} h^* \\ \star^\dagger  & \left(\frac{vw}{v_\varphi}\right)^2 h(f^{K_3})^{-1}\tilde{y}^{d_3*} Y_d^2 (f^{K_3*})^{-1}h^* 
\end{pmatrix},\\ 
M_{{\bf D}^\prime}^2 &=&  \frac{v_\varphi^2}{2}
f^{K_3}f^{K_3*},\nn
\end{eqnarray}
with $(A^d)^2 =h^d h^{d\dagger} +h^{d_3} h^{d_3\dagger}$ and $Y_d^2=\tilde{y}^d \tilde{y}^{d\dagger} + \tilde{y}^{d_3}\tilde{y}^{d_3*}$.
The states associated with the new mass matrices are, respectively,
\bea
d^\prime_{aL} &=& d_{aL} -\frac{vw}{v_\varphi^2}\frac{\left(h^d\tilde{y}^{d\dagger}+h^{d_3}\tilde{y}^{d_3*}\right)_a }{f^{K_3}f^{K_3*}}\mathcal{D}_{2L},  \\ 
d^\prime_{3L} &=& d_{3L} -\frac{v}{v_\varphi}\frac{h}{f^{K_3}}\mathcal{D}_{2L},\nn
\eea
and
\be
\mathcal{D}^\prime_{2L} = \mathcal{D}_{2L} + \frac{vw}{v_\varphi^2}
\frac{\left(\tilde{y}^d h^{d\dagger}+\tilde{y}^{d_3} h^{d_3\dagger}\right)_{a}}{f^{K_3}f^{K_3*}}d_{aL} +\frac{v}{v_\varphi}\frac{h^*}{f^{K_3*}} d_{3L}.
\ee 

In contrast to the up-type quark case, the first two families of the ordinary quarks get masses proportional to $v$, while the third one gets a mass proportional to $(w/v_\varphi)v$. Therefore, in order to get the observed down-type quark masses the Yukawa couplings need to be finely adjusted.

The remaining down-type quarks mix according to the mass matrix $M_{\bf D}^{(2)}$ in Eq.~(\ref{newDmass}). As the off-diagonal terms are much smaller than the diagonal ones, the dominant contributions to the mass eigenvalues are the diagonal terms themselves. Therefore, we have two heavy quarks with masses proportional to $w$ and $v_\varphi$, and small mixing angles. 

Finally, for the quarks with exotic charges, $\mathcal{A}^{(5/3)}_a$ and $\mathcal{B}^{(-4/3)}$, we obtain the following mass matrices:
\bea \label{exocticQmass}
M_{\mathcal{A}} &=& \frac{v_\varphi}{\sqrt{2}}  f^{K_a}_{[2\times2]} \quad \mbox{and}\quad
M_{\mathcal{B}} = \frac{v_\varphi}{\sqrt{2}}  f^{K_3}.
\eea

\section{One-loop effective potential}\label{sec:disc}

In this section, we return to the study of the scalar potential, now at loop level. As explicitly shown in Sec. \ref{scsect}, the CP-even scalar field $S$ in Eq. (\ref{scalon}), defining the flat direction of the potential, remains massless at tree level. In what follows, making use of all the tree-level masses derived in the previous sections for the different sectors of the model, we calculate the one-loop effective potential along the flat direction. Finally, we analyse the stability of the effective potential, and show that the field $S$, the scalon, becomes massive as a result of the breaking of scale invariance at loop level.  

A small curvature in the scalar potential along the radial coordinate, $\phi_r$, is produced when one-loop terms, $V_{\mathrm{1-loop}}$, are included. It implies that the tree-level minimum, $\langle\phi_r\rangle\mathbf{n}$ in Eq. \eqref{nsquared}, picks a definite value $\langle\phi_r\rangle$, and its direction shifts in a $\delta \mathrm{\Phi}$ direction in the field space. In other words, the one-loop minimum turns out to be $\langle\phi_r\rangle\mathbf{n}+\delta \mathrm{\Phi}$. The basic equation determining $\langle\phi_r\rangle$ is 
\begin{equation}
0=\left[ \frac{\partial V_{\mathrm{1-loop}}(\phi_r\mathbf{n})}{\partial\phi_r}\right]_{\langle\phi_r\rangle}. \label{minimum1loop}
\end{equation}
Once $\langle\phi_r\rangle$ is calculated using the previous equation, $\delta \mathrm{\Phi}$ can be found to first order in perturbation theory using
\begin{equation}
0=\textrm{P}_{ij}\delta \mathrm{\Phi}_j\langle\phi_r\rangle^2+\left[ \frac{\partial  V_{\mathrm{1-loop}}(\mathrm{\Phi})}{\partial\mathrm{\Phi}_i}\right]_{\langle\phi_r\rangle \mathbf{n}},
\end{equation}
where $\textrm{P}_{ij}$ is the Hessian matrix in Eq. (\ref{Hessian}). Thus, we must first find $V_{\mathrm{1-loop}}$. In the $\overline{\textrm{MS}}$ renormalisation scheme, this is 
\begin{eqnarray}
V_{\mathrm{1-loop}}(\phi_r\mathbf{n})=A \phi_r^{4}+B\,\phi_r^{4}\ln\left(\frac{\phi_r^{2}}{\mu_0^{2}}\right),
\label{vloop}
\end{eqnarray}
where $\mu_0$ is the same renormalisation scale in Eq. \eqref{lambda7}. Moreover, the dimensionless coefficients $A$ and $B$ are 
\begin{eqnarray}\label{CoeffA}
A=\frac{1}{64\pi^{2}\langle\phi_r\rangle^{4}}&& \left[ \sum_{\mathcal{S}}n_{\mathcal{S}}\,m_{\mathcal{S}}^{4}\left(\ln\frac{m_{\mathcal{S}}^{2}}{\langle\phi_r\rangle^{2}}-\frac{3}{2}\right)+ 3\sum_{\mathcal{V}}n_{\mathcal{V}}\,m_{\mathcal{V}}^4\left(\ln \frac{m_{\mathcal{V}}^{2}}{\langle\phi_r\rangle^2}-\frac{5}{6}\right)\right. \nonumber\\
&&\left.-4\sum_{ \mathcal{F}}\,n_{\mathcal{C}} n_\mathcal{M} \mathrm{Tr}\left[M_{\mathcal{F}}^{4}\left(\ln\frac{M_{\mathcal{F}}^{2}}{\langle\phi_r\rangle^2}-1\right)\right]\right],
\end{eqnarray}
and
\begin{eqnarray}
B&=& \frac{1}{64\pi^{2}\langle\phi_r\rangle^{4}}\left[\sum_{\mathcal{S}}n_{\mathcal{S}}\,m_{\mathcal{S}}^{4}+3\sum_{\mathcal{V}}n_{\mathcal{V}}\,m_{\mathcal{V}}^4-4\sum_{ \mathcal{F}}\,n_{\mathcal{C}} n_\mathcal{M} \mathrm{Tr}\left[M_{\mathcal{F}}^{4}\right]\right],
\label{CoeffB}
\end{eqnarray}
where $\mathcal{S}=H^{\pm}, h, H, A_\varphi$, $\mathcal{V}=W^{\pm},V^{\pm},V^{0(*)},Z_{1,2}$ and $\mathcal{F}=E,{\bf {\tilde E}}, N_2,{\bf \tilde N},{\bf U}^{(1)},{\bf U}^{(2)},{\bf D}^{(1)},{\bf D}^{(2)},{\bf \mathcal{A}},{\bf \mathcal{B}}$. We also have that $m_\mathcal{S},m_\mathcal{V}$ are the tree-level masses of the scalars and vector bosons, respectively, as given in Eqs. (\ref{ChVm}), (\ref{Zsmass}), (\ref{masahpm}), (\ref{h1h2}) and (\ref{massA}). Similarly, $M_{\mathcal{F}}$ represents the mass matrices of the fermions, leptons and quarks, given in Eqs. (\ref{Emass}), (\ref{Etilmass}), (\ref{N2mass}), (\ref{NtilM}),  (\ref{newUmass}), (\ref{newDmass}) and (\ref{exocticQmass}). Furthermore,  $n_{\mathcal{S,\,V}}=2$ for $\mathcal{S}=h^{\pm}$ and $\mathcal{V}=W^{\pm},V^{\pm},V^{0(*)}$ and equal to $1$ otherwise. $n_{\mathcal{C}}=3$ for $\mathcal{F}={\bf U}^{(1)},{\bf U}^{(2)},{\bf D}^{(1)},{\bf D}^{(2)},{\bf \mathcal{A}},{\bf \mathcal{B}}$ and equal to $1$ otherwise. Finally, $n_\mathcal{M} =1/2$ for Majorana fermions, and $1$ otherwise.

After obtaining $V_{\mathrm{1-loop}}(\phi_r\mathbf{n})$, we can use Eq. \eqref{minimum1loop} to find
\begin{eqnarray}
\langle\phi_r\rangle=\mu_0\exp{\left[-\frac{1}{4}-\frac{A}{2B}\right]},
\label{mini}
\end{eqnarray}
showing that the scale of the symmetry-breaking parameter $\langle\phi_r\rangle$ is set by the renormalisation scale $\mu_0$. Now, we can use Eq. (\ref{mini}) to eliminate the explicit dependence of the effective potential, $V_{\mathrm{1-loop}}(\phi_r\mathbf{n})$, on the renormalisation scale $\mu_0$,  {\it i.e.}
\be
V_{\mathrm{1-loop}}(\phi_r\mathbf{n})=B\,\phi_r^{4}\left[\ln\left(\frac{\phi_r^{2}}{\langle \phi_r\rangle^{2}}\right)-\frac{1}{2}\right],
\label{vloop2}
\ee
which is valid for $B\neq0$. It is important to realise that the stationary point, $\langle\phi_r\rangle\mathbf{n}$, is not a minimum unless $B>0$, because $V_{\mathrm{1-loop}}$ is not bounded from below if $B<0$. Note that in the case of $B=0$, the scalar potential is purely quartic, {\it cf.} Eq. \eqref{vloop}. Therefore, as can be seen from Eq. (\ref{CoeffB}), the $B>0$ condition imposes a constraint on the masses of the particles in the model. More specifically, the fermion masses must not dominate since they contribute negatively to $B$.

Additionally, as a consequence of the scale-invariance breaking, the following scalon mass is obtained from the effective potential in Eq. (\ref{vloop2}):
\begin{eqnarray}
m_{S}^{2}=8B\langle\phi_r\rangle^2,
\label{1loopmass}
\end{eqnarray}
which is positive for a bounded-from-below potential since, in this case, $B>0$.

To determine the condition for the stability of the effective potential, let us estimate $B$ by taking into account the vev hierarchy used throughout this paper, {\it i.e.} $v\ll w \ll v_\varphi (\simeq\langle \phi_r\rangle)$. Within this hierarchy, we can neglect, at leading order, contributions coming from particles with masses around the scales $v$ and $w$, such as all of those coming from the vector bosons. Thus, the dominant contributions to $B$ come from the heaviest particles in the model and can be written as
\begin{equation}
\label{bvarphi}
B\simeq \frac{1}{64\pi^{2}v_\varphi^4}\left[ m_{A_\varphi}^4 - 4\mathrm{Tr}\left[M_\mathcal{E^\prime}^4 + M_{N_2}^4 + \frac{1}{2}M_{{\bf N}^\prime}^4+3\left(\sum_{i=1}^{2}\left( M_{\bf U}^{(i)\,4}+M_{\bf D}^{ (i)\,4}\right) +M_{\mathcal{A}}^{4}+M_{\mathcal{B}}^{4}\right)\right]\right],
\end{equation}
where the scalar field mass is given in Eq. (\ref{massA}), the lepton masses come from Eqs. (\ref{clMasses}), (\ref{N2mass}) and (\ref{numass}), and the quark masses can be obtained from Eqs. (\ref{newUmass}), (\ref{newDmass}) and (\ref{exocticQmass}).

From Eq. (\ref{bvarphi}), we see that the potential stability at one-loop level can be determined by the interplay between the heavy masses of the pseudoscalar $A_\varphi$ and the extra fermions in the model. In order for $B$ to be positive, the pseudoscalar mass, $m_{A_\varphi}$, must be large enough to compensate for the negative contributions coming from several heavy fields in the fermion sector. To see how this can be achieved without resorting to unnatural assumptions, we consider a simple scenario where the Yukawa couplings associated with the fermion masses proportional to $v_\varphi$ in Eq. (\ref{bvarphi}) are of order one. In this case, we obtain
\begin{eqnarray}
B&\simeq& \frac{1}{64\pi^{2}}\left(64|\lambda_\varphi^\prime|^{2}-\frac{75}{2}\right),
\end{eqnarray}
which is positive for $|\lambda_\varphi^\prime|\gtrsim 0.77$, a value still well within the perturbative region. For such coupling constant values, the heavy fermions and pseudoscalar $A_\varphi$ have masses around $v_\varphi=10^3$ TeV and therefore lie outside the energy range of current and near-future colliders. Similarly, for $|\lambda_\varphi^\prime| \simeq1$, the scalon mass is $m_S\simeq 580$ TeV, which is also too large to be produced at colliders in the foreseeable future. Therefore, in this scenario, all the fields added to the model in Sec. \ref{sec:ourmod}, {\it i.e.} the scalar singlet and the vector-like fermions, which play a crucial role in the generation of SM fermion masses as well as in the consistent breaking of scale invariance, can be integrated out. The resulting effective theory contains only the same degrees of freedom as the 3-3-1 model with two scalar triplets shown in Sec. \ref{sec:minmod}. However, contrary to what we have seen in Sec. \ref{sec:minmod}, all particles are now massive as required by experimental evidence.

\section{Residual symmetry and phenomenological implications}\label{sec:resS}

In addition to the conservation of the baryon number, $U(1)_B$, the minimal scale-invariant 3-3-1 model presents another accidental global symmetry, $U(1)_N$, which follows from the imposition of the $Z_8$ discrete symmetry. Although the $U(1)_N$ is spontaneously broken when the scalar triplets acquire vevs, a residual symmetry $U(1)_{\mathcal{G}}$, generated by
\be 
\mathcal{G} = \,-4T_3 +2\sqrt{3} T_8+N\,,
\ee
where $N$ represents the $U(1)_N$ charge, remains exactly conserved. In Table \ref{tab2}, we show how the fields transform under $U(1)_N$ and $U(1)_{\mathcal{G}}$.
\begin{table}[h!]
    \centering
\begin{tabular}{|c||c|c|c|c|c|c|c|c|c|c|c|c||c|c||c|c|c|}
        \hline
  &  $\psi_{iL}$ & 
 $ e_{i R}$  &
 $ E_{i R}$   &
 $Q_{a L}$  &  $ Q_{3L}$  &
 $u_{iR}$  &
 $U_{aR}$  &
 $d_{iR}$  &  $D_{R}$  &
 $\Psi_{iL,R}$  &
 $K_{a L,R}$  &
 $ K_{3L,R}$  &
 $\rho$  &
 $\chi$ &
 $W^+_\mu$  &
 $V^+_\mu$  &
 $V^0_\mu$ \\
 \hline\hline
$U(1)_N$ & $1$ & $4$ & $-1$ & $2$ & $0$ & $-1$ & $4$ & $3$ & $-2$ & $-3$ & $1$ & $1$ & $1$ & $2$ & $0$ & $0$ & $0$ \\
\hline
$U(1)_{\mathcal{G}}$ & $\begin{pmatrix}
0 \\ 4 \\ -1
\end{pmatrix}$ & $4$ & $-1$ & $\begin{pmatrix}
3 \\ -1 \\ 4
\end{pmatrix}$ &   $\begin{pmatrix}
-1 \\ 3 \\ -2
\end{pmatrix}$ & $-1$ & $4$ & $3$ & $-2$ &   $\begin{pmatrix}
-4 \\ 0 \\  -5
\end{pmatrix}$ &  $\begin{pmatrix}
0 \\ 4 \\ -1
\end{pmatrix}$ &  $\begin{pmatrix}
2 \\ -2 \\ 3
\end{pmatrix}$ &  $\begin{pmatrix}
0 \\ 4 \\ -1
\end{pmatrix}$ &  $\begin{pmatrix}
1 \\ 5 \\ 0
\end{pmatrix}$ & $-4$ & $1$ & $5$\\
\hline
\end{tabular} 
\caption{Field charges under the $U(1)_N$ and $U(1)_{\mathcal{G}}$ symmetries. The fields not shown above do not carry charges under these symmetries.}
    \label{tab2}
\end{table}

Let us point out two important differences between $U(1)_{\mathcal{G}}$ and the residual symmetry present in the 3-3-1 model with two Higgs triplets, discussed in Sec. \ref{sec:minmod}. First, contrary to what happens in the model in Sec. \ref{sec:minmod}, neither $U(1)_N$ nor $U(1)_{\mathcal{G}}$ are Peccei-Quinn-like symmetries since the associated $[SU(3)_C]^2\otimes U(1)_N$ anomaly coefficient vanishes identically. Second, $U(1)_{\mathcal{G}}$ is not chiral with respect to any left-handed fermion triplet component and its right-handed singlet counterpart; thus, as shown in Sec. \ref{sec:fermass}, all fermions become massive.

Another distinctive feature arising from the residual $U(1)_{\mathcal{G}}$ symmetry is the stabilisation of the lightest among the new particles that do not mix with the SM ones. This can be more easily understood by considering the linear combination of the two conserved global symmetries generated by $\mathcal{G}^\prime = \mathcal{G}-3\mathbf{B}$, where $\mathbf{B}$ is the field's baryon number. It is straightforward to see that the symmetry generated by $\mathcal{G}^\prime$, $U(1)_{\mathcal{G}^\prime}$, is conserved and so is its parity subgroup defined by $\mathcal{P}=(-1)^{\mathcal{G}^\prime}$.  In Table \ref{tab3}, we show how the fields transform under $U(1)_{\mathcal{G}^\prime}$ and $\mathcal{P}$. We see that all the SM fields transform trivially under $\mathcal{P}$. Consequently, the lightest amongst the $\mathcal{P}$-odd fields cannot decay into SM particles and is stable. A parity symmetry resembling the one obtained here has been observed and explored in the context of dark matter stability in different 3-3-1 realisations \cite{Dong:2013wca,Dong:2014wsa,Dong:2015yra, Alves:2016fqe, Leite:2019grf}. 

\begin{table}[h!]
    \centering
\begin{tabular}{|c||c|c|c|c|c|c|c|c|c|c|c|c||c|c||c|c|c|}
        \hline
  &  $\psi_{iL}$ & 
 $ e_{i R}$  &
 $ E_{i R}$   &
 $Q_{a L}$  &  $ Q_{3L}$  &
 $u_{iR}$  &
 $U_{aR}$  &
 $d_{iR}$  &  $D_{R}$  &
 $\Psi_{iL,R}$  &
 $K_{a L,R}$  &
 $ K_{3L,R}$  &
 $\rho$  &
 $\chi$ &
 $W^+_\mu$  &
 $V^+_\mu$  &
 $V^0_\mu$ \\
 \hline
\hline
$U(1)_{\mathcal{G}^\prime}$ & $\begin{pmatrix}
0 \\ 4 \\ -1
\end{pmatrix}$ & $4$ & $-1$ & $\begin{pmatrix}
2 \\ -2 \\ 3
\end{pmatrix}$ &   $\begin{pmatrix}
-2 \\ 2 \\ -3
\end{pmatrix}$ & $-2$ & $3$ & $2$ & $-3$ &   $\begin{pmatrix}
-4 \\ 0 \\  -5
\end{pmatrix}$ &  $\begin{pmatrix}
-1 \\ 3 \\ -2
\end{pmatrix}$ &  $\begin{pmatrix}
1 \\ -3 \\ 2
\end{pmatrix}$ &  $\begin{pmatrix}
0 \\ 4 \\ -1
\end{pmatrix}$ &  $\begin{pmatrix}
1 \\ 5 \\ 0
\end{pmatrix}$ & $-4$ & $1$ & $5$\\
\hline
$\mathcal{P}$ & $\begin{pmatrix}
+ \\ + \\ -
\end{pmatrix}$ & $+$ & $-$ & $\begin{pmatrix}
+ \\ + \\ -
\end{pmatrix}$ &   $\begin{pmatrix}
+ \\ + \\ -
\end{pmatrix}$ & $+$ & $-$ & $+$ & $-$ &   $\begin{pmatrix}
+ \\ + \\  -
\end{pmatrix}$ &  $\begin{pmatrix}
- \\ - \\ +
\end{pmatrix}$ &  $\begin{pmatrix}
- \\ - \\ +
\end{pmatrix}$ &  $\begin{pmatrix}
+ \\ + \\ -
\end{pmatrix}$ &  $\begin{pmatrix}
- \\ - \\ +
\end{pmatrix}$ & $+$ & $-$ & $-$\\
\hline
\end{tabular} 
 \caption{Field charges under $U(1)_{\mathcal{G}^\prime}$ and its parity subgroup $\mathcal{P}=(-1)^{\mathcal{G}^\prime}$. The fields not displayed here transform trivially under these symmetries.}
    \label{tab3}
\end{table}

If the lightest parity-odd field is electrically neutral, as it is the case of $N_{2}$ and $V^0$, it can play the role of a stable dark matter candidate\footnote{Notice that $\chi_2^0$, the only parity-odd neutral scalar, is the would-be Goldstone boson absorbed by $V^0$ and should not be considered in this analysis.}. As shown in the previous sections, the assumed vev hierarchy implies that $m_{N_2}(v_\varphi) \gg m_{V^0}(w)$, so that the complex neutral vector field $V^0$ is the lightest $\mathcal{P}$-odd field. Despite its stability, the vector boson $V^{0}$ could only compose a small fraction of the dark matter in the Universe, as pointed out in Refs.~\cite{Mizukoshi:2010ky,Alvares:2012qv,Dong:2013wca} for a different model but which contains the same $V^{\pm}$ and $V^{0}$ vector bosons. Nonetheless, $V^0$ appears as missing energy in the production process signals of the new heavy fermions, as we comment in what follows.  

At this point, it is important to note the expected signals of the new fermions production predicted at the TeV scale. Due to the hierarchy of the vevs, the new fermions that could be first observed are those whose masses are directly proportional to the scale $w$. These are the two $U_{a}$ quarks (which mix with the ${\cal U}_{1a}$ quarks), the $D$ quark (which mix with the ${\cal D}_{1}$ quark) and the heavy $E_i$ leptons, whose masses are given, respectively, by Eqs. (\ref{newUmass}), (\ref{newDmass}) and  (\ref{Emass}). Such fermions carry non-trivial charges under the $U(1)_{\mathcal{G}^\prime}$ symmetry and are odd under the parity $\mathcal{P}$, as shown in Table \ref{tab3}, implying that they can only be produced in pairs. Also, these fermions cannot decay into a final state containing only SM particles, since all SM particles are $\mathcal{P}$ even. 
Being the neutral complex gauge field $V^0$ the lightest  $\mathcal{P}$-odd  particle, the production of the new fermions has a signature of final states with SM particles plus missing energy. 

The model has then some characteristic signals that could be studied at colliders. Let us assume that the $D$ quark is the lightest $\mathcal{P}$-odd fermion and that its main decay modes are those involving the gauge interactions, $D\rightarrow b\,V^0$ and $D\rightarrow t\,V^-$. Then, the pair production of the $D$ quark would lead to the following final states:
\begin{eqnarray}
    p\,p\longrightarrow D\,\overline{D}\, & &\longrightarrow b\,V^0\,\,\,\overline{b}\,V^{0\dagger},\nonumber\\
     & &\longrightarrow b\,V^0\,\,\,\overline{t}\,V^{+}\longrightarrow b\,V^0\,\,\,\overline{t}\,{t}\,\overline{b}\,V^{0\dagger},\nonumber\\
     & &\longrightarrow  t\,V^-\,\,\,\overline{b}\,V^{0\dagger}\longrightarrow t\,\overline{t}\,b\,V^0\,\,\,\overline{b}\,V^{0\dagger},\nonumber\\
    & &\longrightarrow t\,V^-\,\,\,\overline{t}\,V^{+}
    \longrightarrow t\,\overline{t}\,b\,V^0\,\,\,\overline{t}\,{t}\,\overline{b}\,V^{0\dagger},
    \label{Ddecay}
\end{eqnarray}
with the decay modes  $V^-(V^+)\rightarrow \overline{t}\,D^*({t}\,\overline{D}^*)\rightarrow \overline{t}\,b\,V^0({t}\,\overline{b}\,V^{0\dagger})$, where $D^*(\overline{D}^*)$ is a virtual intermediary state. Considering the $SU(3)_L\otimes U(1)_X$ symmetry breaking scale being  $w\simeq 10$ TeV, as in Sec. \ref{sec:fermass} for the SM fermion mass generation mechanism, assuming $g\simeq 0.653$, we have $m_{V^0}\simeq 3.264$ TeV for the $V^0$ mass and, therefore, $m_D\geq m_{V^0}+m_b>3.264$ TeV. The first production signal in Eq.~(\ref{Ddecay}) gives two $b$-$jets$ plus missing energy in the form of $V^0,\,V^{0\dagger}$ vector bosons. Such a signal is similar to the one in the searches of the bottom squark pair production, with the missing energy carried by the lightest supersymmetric particle (the neutralino), that has been investigated by the CMS and ATLAS Collaborations within the contest of simplified models~\cite{Sirunyan:2017kiw,Aaboud:2017phn}. But the limits resulting from these experiments for the masses of the bottom squark and the lightest supersymmetric particle are well below the $D$ quark and $V^0$ masses we are considering here and, therefore, cannot be used to constrain the model. The remaining three production signals in Eq.~(\ref{Ddecay}) would be more difficult to observe because they involve more than two $b$-$jets$, once $t\rightarrow b\,W^{+}$, plus decays from $W^\pm$ bosons. The production of $V^0$ in the next LHC run is, however, very unlikely due to the assumed value for $m_{V^0}$.  Nevertheless,  direct signals of the new fermions and vector bosons, whose masses are directly related to the scale $w$, {\it i.e.} $E_i$, $U_a$, $D$ and $V^\pm$, $V^0$, $Z_2$, could show up in the high-luminosity LHC and the projected future circular collider. It would be worthwhile to perform a dedicated study of such signals, but this is outside the scope of this work. 

\begin{table}[h!]
    \centering
\begin{tabular}{|c||cc|ccc|c|}
        \hline
  &\,\, $H$ \,\, & \,\, $H^\pm$ \,\, &  \,\,
 $V^0_\mu$ \,\, &
\,\, $V^\pm_\mu$\,\,  &
\,\, $Z_{2\mu}$ \,\, & \,\,$ E_{i},U_{a},D$\,\,\\
 \hline
\hline
Mass (GeV) & 800  &  3500  &  3264 &
 3265 &
 3974  &
 3535\\
\hline
\end{tabular} 
 \caption{Mass benchmarks for the particles at the 3-3-1 scale $w=10$ TeV. See the text for details. }
    \label{tab4}
\end{table}

As observed in Sec.~\ref{scsect}, the spectrum of the scalar particles at the $w$ scale is composed of the CP-even $H$, which has a small mixing with the 125 GeV Higgs boson $h$, and the charged scalar $H^\pm$. Their approximate masses, $m_{H}\approx \sqrt{2\lambda_{\chi}} w$ and $m_{H^\pm}\approx\sqrt{\frac{\lambda_{\rho\chi}^\prime}{2}}w$, as given in Eqs. (\ref{mh12}) and (\ref{masahpm}), depend on the free parameters $\lambda_{\chi}$ and $\lambda_{\rho\chi}^\prime$ of the scalar potential. The charged scalar $H^\pm$, defined in Eq.~(\ref{Hpm}), is a linear combination of $\rho_3^\pm$ and $\chi_1^\pm$ and therefore, according to Table \ref{tab3}, is odd under $\mathcal{P}$. Consequently, $H^\pm$  must be heavier than the assumed lightest  $\mathcal{P}$-odd particle $V^0$. The neutral scalar $H$, on the other hand, is $\mathcal{P}$ even and, as such, it might be the lightest particle at the scale $w$. Such a particle could be produced in the next run of the LHC. Due to the hierarchy among the vevs, $H$ couples mostly to the new fermions $E_i$, $U_a$ and $D$, as seen from Eqs.~(\ref{Yuklep}) and (\ref{Yukq}). Thus, in a proton-proton collider, the principal production channel of $H$ would be through gluon fusion, as it happens in the case of the 125-GeV Higgs boson, with the production cross section given by the sum of terms directly proportional to the Yukawa couplings $y_{aa}^U$ and $y_{34}^D$ in the operators $\frac{y_{aa}^U}{\sqrt2} H\overline{U_a}U_a$ and $\frac{y_{34}^D}{\sqrt2} H\overline{D}D$. Although it is not the aim of the present work, it would be interesting to perform a phenomenological study of the distinct signals of heavy Higgs boson H along with the other predicted particles at the scale $w$. At last, we present in Table \ref{tab4} a benchmark for the particle spectrum at the assumed scale $w=10$ TeV. For the scalars, we obtain  $m_{H}= 800$ GeV and $m_{H^\pm}= 3.5$ TeV by assuming that $\lambda_{\chi}=3.2\times10^{-3}$ and $\lambda_{\rho\chi}^\prime=0.245$, whereas for the fermion masses of $3.535$ TeV, we took $y^E=y^U=y^D\simeq0.5$. Notice that, despite being $\mathcal{P}$ even, it follows from Eq. (\ref{mrel}) that the neutral vector boson $Z_2$ is necessarily heavier than the lightest $\mathcal{P}$-odd field $V^0$. 

\section{Conclusions}\label{sec:conc}

In this paper, we have proposed the minimal scale-invariant 3-3-1 model, based on the $\331$ gauge symmetry and scale invariance. It extends the effective 3-3-1 model with two scalar triplets \cite{Barreto:2017xix}, reviewed in Sec. \ref{sec:minmod}, which, despite the attractiveness of a very compact scalar spectrum, is not phenomenologically viable. The issue being the existence of an accidental chiral symmetry that forbids some of the standard fermions to become massive. To generate tree-level masses to all the fermions, in Sec. \ref{sec:ourmod}, we have introduced vector-like quark and lepton triplets and lepton singlets, the latter necessary for the generation of neutrino masses. Furthermore, the scalar sector is kept as minimal as possible with only an extra singlet being added to allow for consistent mechanisms of dynamical symmetry breaking and fermion mass generation. 

The study of the fermion spectrum in Sec. \ref{sec:fermass} has shown that, with the inclusion of the extra fermions, no accidental chiral symmetry remains present, and all the fermions become massive. This is easier to see with the use of the $Z_8$ symmetry in Table \ref{tab1}, which has at least two important roles. First, it greatly simplifies the Yukawa and scalar Lagrangians. Second, together with the gauge and scale symmetries, the $Z_8$ symmetry makes evident the seesaw texture in most of the fermion mass matrices provided that $v_\varphi \gg w \gg v$, where $v_\varphi$ is the scale associated with the scalar singlet, $w$ is the 3-3-1 breaking scale, and $v$ is the electroweak scale. This point is useful to mitigate possible phenomenological issues associated with flavour changing neutral currents because, in this case, the suppressed  mixing between light and heavy fermions are proportional to $v/v_\varphi$, $v\,w/v^2_\varphi$ or $w/v_\varphi$. Thus, if, for instance, $w=10$ TeV as expected for the 3-3-1 models, then $v_\varphi=10^3$ TeV largely reduces such undesirable phenomena without resorting to fine-tuning on the parameters of the model.

Interestingly, once the seesaw mechanism takes place, the heavy masses of the extra fermions, proportional to $v_\varphi$, suppress the masses of some of the standard ones.  For instance, the first two families of up-type quarks get seesaw suppressed masses $\propto (w/v_\varphi)v$, while the third family gets a mass proportional to the electroweak scale $v$, providing thus an explanation for the mass hierarchy between the third and first two families when assuming, {\it e.g.}, that $v_\varphi= 10^3$ TeV and $w=10$ TeV as previously. Similarly, charged leptons get seesaw suppressed masses $\propto (w/v_\varphi)v$ suggesting an origin for the hierarchy between their masses and the electroweak scale.

The minimal scalar sector, containing two triplets and one singlet, is one of the most appealing features of the proposed model. In Sec. \ref{scsect}, we have derived the analytical conditions at tree level that must be satisfied by a bounded-from-below potential, {\it cf.} Eq. \eqref{cop_1}, using the copositivity method. We have seen that such conditions are automatically satisfied when the Hessian matrix,  Eq. \eqref{semipositive}, and the scalar masses, Eq.  \eqref{masahpm}, are positive semidefinite. Moreover, we have shown that the potential exhibits a flat direction, which defines the scalon field $S$. To assess the consistency of the dynamical symmetry breaking {\it \`a la} Coleman-Weinberg, in Sec. \ref{sec:disc}, we have calculated the one-loop effective potential along the flat direction using the Gildener-Weinberg method. In the limit $v_\varphi\gg w \gg v$, we have shown that the stability of the effective potential is basically determined by the interplay between the masses of the pseudo-scalar $A_\varphi$ and the heavy extra fermions. We have found that the potential stability can be naturally assured for couplings of order 1. In such a case, since $v_\varphi=10^3$ TeV, the scalon, with a mass of $\approx 580$ TeV, the CP-odd $A_\varphi$ and the new fermions with masses $\propto v_\varphi$ are too heavy to be produced at current or near-future experiments.
Nonetheless, the 3-3-1 fields with masses proportional to $w=10$ TeV -- all the non-SM vector bosons, the scalars $H$ and $H^\pm$, and the fermions $E_i$, $U_a$ and $D$ -- could be produced at the LHC and the Future Circular Collider. Most of these fields, with the exception of $H$ and $Z^\prime$, are odd under the residual parity symmetry $\mathcal{P}$ and as such cannot decay into the $\mathcal{P}$-even SM particles only. Thus, an expected signature following the production and subsequent decay of the $\mathcal{P}$-odd particles in colliders is the presence of the complex neutral vector boson $V^0$, the lightest $\mathcal{P}$-odd particle, as missing energy.

\acknowledgments
A. G. Dias thanks Conselho Nacional de Desenvolvimento Cient\'{\i}fico e Tecnol\'ogico (CNPq) for its financial support under the grant 305802/2019-4. J. Leite acknowledges financial support under grants 2017/23027-2 and 2019/04195-7, S\~ao Paulo Research Foundation (FAPESP). At IFIC, J. Leite was also supported by the Spanish grants FPA2017-85216-P (AEI/FEDER, UE), PROMETEO/2018/165 (Generalitat Valenciana) and the Spanish Red Consolider MultiDark FPA2017-90566-REDC. This study was also financed in part by the Coordena\c{c}\~{a}o de Aperfei\c{c}oamento de Pessoal de N\'ivel Superior - Brasil (CAPES) - Finance Code 001 (W.C. Vieira). B. L. S\'anchez-Vega would like to thank Pr\'o-Reitoria de Pesquisa da Universidade Federal de Minas Gerais for its financial support and UFABC for kind hospitality.

\bibliographystyle{apsrev4-1}

\bibliography{3-3-1}

\end{document}